\documentclass[12pt]{article}
\usepackage{latexsym}
\usepackage{cite,epsfig,amssymb,euscript,slashed}
\usepackage{amsmath}
\usepackage{array,calc,epsfig}
\usepackage{bbm}
\usepackage{fancybox}

\def\be{\begin{equation}}
\def\ee{\end{equation}}
\def\bseq{\begin{subequations}}
\def\eseq{\end{subequations}}

\def\bea{\begin{eqnarray}}
\def\eea{\end{eqnarray}}

\def\bseq{\begin{subequations}}
\def\eseq{\end{subequations}}

\numberwithin{equation}{section} 
\usepackage[]{graphicx}

\def\d {{\rm d}}

\def\ri{{\rm i}}
\def\re{{\rm e}}
\def\rd{{\rm d}}

\def\cald         {{\cal D}}

\def\calf         {{\cal F}}
\def\calg         {{\cal G}}

\def\calm         {{\cal M}}
\def\caln         {{\cal N}}

\newcommand {\cC}{{\cal C}}
\newcommand {\cD}{{\cal D}}

\newcommand {\cF}{{\cal F}}

\newcommand {\cN}{{\cal N}}

\newcommand {\cS}{{\cal S}}

\newcommand {\cV}{{\cal V}}
\newcommand {\cW}{{\cal W}}

\def\del          {\partial}

\def\ii           {{\rm i}}

\def\sqr#1#2{{\vcenter{\vbox{\hrule height.#2pt
 \hbox{\vrule width.#2pt height#1pt \kern#1pt \vrule width.#2pt}\hrule
 height.#2pt}}}}

\def\a{\alpha}
\def\b{\beta}

\def\g{\gamma}
\def\G{\Gamma}

\def\j{\psi}
\def\k{\kappa}
\def\l{\lambda}

\def\q{\theta}

\def\s{\sigma}

\def\x{\xi}
\def\z{\zeta}

\def\J{\Psi}
\def\L{\Lambda}

\def\S{\Sigma}
\def\U{\Upsilon}
\def\X{\Xi}

\def\g{\gamma}
\def\l{\lambda}

\def\d{\text{d}}

\newcommand{\gd}{{\dot\g}}

\newcommand{\ad}{{\dot{\alpha}}}
\newcommand{\bd}{{\dot{\beta}}}

\newcommand{\ve}{\varepsilon}

\newcommand{\pa}{\partial}
\newcommand{\hf}{\frac12}

\def\slashchar#1{\setbox0=\hbox{$#1$}           
\dimen0=\wd0                                 
\setbox1=\hbox{/} \dimen1=\wd1               
\ifdim\dimen0>\dimen1                        
\rlap{\hbox to \dimen0{\hfil/\hfil}}      
#1                                        
\else                                        
\rlap{\hbox to \dimen1{\hfil$#1$\hfil}}   
/                                         
\fi}

\begin{document}
\font\cmss=cmss10 \font\cmsss=cmss10 at 7pt



\title{The Goldstino Brane, the Constrained Superfields and Matter in  $\mathcal N=1$ Supergravity}

\date{}

\maketitle

\vspace{-1.5cm}

\begin{center}
{\textsl{\rm Igor Bandos$\,^{a,b}$,  Markus Heller$^{c,d}$, Sergei M. Kuzenko$\,^f$,\\ Luca Martucci$\,^{c,g}$ and Dmitri Sorokin$\,^{g,c}$ }}

\vspace{0.5cm}
\textit{\small $^a$ Department of Theoretical Physics, University of the Basque Country UPV/EHU, \\ P.O. Box 644, 48080 Bilbao, Spain}\\ \vspace{6pt}
\textit{\small $^b$  IKERBASQUE, Basque Foundation for Science, 48011, Bilbao, Spain}\\ \vspace{6pt}
\textit{\small $^c$  Dipartimento di Fisica e Astronomia ``Galileo Galilei",  Universit\`a degli Studi di Padova, \\ Via Marzolo 8, 35131 Padova, Italy} \\ \vspace{6pt}
\textit{\small ${}^{d}$ Institut f\"ur Theoretische Physik, Ruprecht-Karls-Universit\"at,
Philosophenweg 19, 69120 Heidelberg, Germany}\\
\vspace{6pt}
\textit{\small ${}^{f}$ School of Physics M013, The University of Western Australia\\
35 Stirling Highway, Crawley W.A. 6009, Australia}\\ \vspace{6pt}
 \textit{\small $^g$ I.N.F.N. Sezione di Padova,
Via Marzolo 8, 35131 Padova, Italy}
\end{center}

\vspace{5pt}

\abstract{

\noindent We show that different (brane and constrained superfield) descriptions for the Volkov-Akulov goldstino coupled to $\mathcal N=1$, $D=4$ supergravity with matter produce similar wide classes of models with spontaneously broken local supersymmetry and discuss the relation between the different formulations.
As with the formulations with irreducible constrained superfields, the geometric goldstino brane approach has the advantage of being manifestly off-shell supersymmetric without the need to introduce auxiliary fields. It provides an explicit solution of the nilpotent superfield constraints and avoids issues with non-Gaussian integration of auxiliary fields. We describe general couplings of the supersymmetry breaking sector, including the goldstino and other non-supersymmetric matter, to supergravity and matter supermultiplets. Among various examples, we discuss a goldstino brane contribution to the gravitino mass term and the supersymmetrization of the anti-D3-brane contribution to the effective theory of type IIB warped flux compactifications.

\bigskip
\bigskip
\centerline{\it Dedicated to the memory of Mario Tonin}

}


\thispagestyle{empty}


\newpage

\setcounter{footnote}{0}

\tableofcontents
\newpage

\section{Introduction}
As is well known, in field theory spontaneous breaking of rigid supersymmetry manifests itself in the presence of massless fermionic spinor fields, Volkov-Akulov goldstini \cite{Volkov:1972jx,Volkov:1973ix,Akulov:1974xz}. In supergravity the goldstini generate a positive contribution to the cosmological constant and, upon having being ``eaten" by gravitini, provide a mass to the latter \cite{Volkov:1973jd,Volkov:1974ai}.  Mechanisms realizing supersymmetry breaking effects in  globally and locally supersymmetric models have been an important subject of intensive research since the very discovery of supersymmetry and supergravity \cite{Volkov:1972jx,Fayet:1974jb,O'Raifeartaigh:1975pr,Volkov:1973jd,Volkov:1974ai,Freedman:1976aw,Deser:1977uq}. Their understanding is indispensable for the construction of phenomenologically relevant supersymmetric models of fundamental interactions and cosmology in which the role of the goldstini has recently undergone a thorough reconsideration, see, e.g.,  \cite{Antoniadis:2014oya,Buchmuller:2014pla,Ferrara:2014kva,Kallosh:2014via,Dall'Agata:2014oka,Kallosh:2014hxa,Dudas:2016eej}
and references therein.

There are two approaches to describe goldstini and their couplings to other fields in the theory. The first one, used in the original papers by Volkov and Akulov \cite{Volkov:1972jx,Volkov:1973ix,Akulov:1974xz}, is the geometrical method of non-linear realizations of spontaneously broken symmetries \cite{Coleman:1969sm,Callan:1969sn,Volkov:1973vd}. This formulation is directly related to mechanisms of spontaneous symmetry breaking caused by extended dynamical objects such as branes in string theory \cite{Hughes:1986dn,Kallosh:1997aw}.
The second approach, in which the goldstini are considered as components of constrained superfields \cite{Ivanov:1977my,Ivanov:1978mx,Rocek:1978nb,Lindstrom:1979kq,Samuel:1982uh,Casalbuoni:1988xh,Komargodski:2009rz,Kuzenko:2011ti}, is more related to conventional superfield constructions of supersymmetric theories, since \emph{a priori} the superfields transform linearly under supersymmetry but due to constraints their independent components, including the goldstino, transform non-linearly. The emergence of the constraints may also be viewed as an effective field theory limit in which certain mass parameters become very large and the corresponding modes decouple. The constrained superfield description has been used in the most of recent literature on spontaneous supersymmetry breaking in supergravity (see e.g. \cite{Kuzenko:2011ti,Farakos:2013ih,Dudas:2015eha,Bergshoeff:2015tra,Hasegawa:2015bza,Ferrara:2015gta,Kuzenko:2015yxa,Antoniadis:2015ala,Kallosh:2015tea,Kallosh:2015sea,Dall'Agata:2015zla,Schillo:2015,Kallosh:2015pho,Ferrara:2016een,Farakos:2016hly} and references therein).

In \cite{Bandos:2015xnf} it has been shown how the Volkov-Akulov ideas have a natural  brane incarnation in a locally supersymmetric context. In this approach one introduces a (space-filling) brane supporting the goldstino, henceforth dubbed {\em goldstino brane}, which couples in a manifestly locally supersymmetric way to the `bulk' supergeometry, i.e.\ to the gravity multiplet.

One of the main purposes of this paper is to illustrate how the goldstino brane allows one to easily  couple the goldstino  to arbitrary matter as well.   As we will discuss in detail, one can  couple the goldstino to `bulk' supersymmetric matter or `non(linearly)-supersymmetric' matter\footnote{We call matter fields non-supersymmetric if they do not form a linear supermultiplet, i.e. supersymmetry is realised non-linearly on such fields.} propagating on the goldstino brane itself. In particular, we will show how in this geometric framework one can naturally generate a supersymmetry breaking contribution to the gravitino mass, while the construction of such a term with the constrained superfields requires the use of a so called special minimal (3-form) formulation \cite{Grisaru:1981xm,Ovrut:1997ur} of
$\mathcal N=1$ supergravity whose coupling to matter is more restricted.

Let us stress that the goldstino brane may have a more fundamental origin, as in some stringy constructions with anti-D-branes in flux compactifications, but it may also be interpreted  as  an auxiliary geometric object, if the microscopic origin of the supersymmetry breaking  has nothing to do with branes in a higher-dimensional theory.

Another main purpose of this paper is to clarify the inter-relation between the goldstino brane approach and the formulations that use different constrained superfields.  In the rigid supersymmetry case, the similar question on the equivalence between the different descriptions of the goldstino  in the absence of matter  was addressed in  \cite{Ivanov:1977my,Ivanov:1978mx,Rocek:1978nb,Lindstrom:1979kq,Samuel:1982uh}  and the explicit form of the non-linear field redefinitions  relating the different formulations were derived in \cite{Kuzenko:2010ef,Kuzenko:2011tj}.
We will extend these results to a more general framework of matter coupled supergravity and show that all these formulations are equivalent to each other (modulo the example of the gravitino mass term of Section \ref{gm}) and describe similar general couplings of the goldstino to supergravity and matter, but one or another of them, like the goldstino brane, may be more suitable for the construction of specific effective models with spontaneously broken supersymmetry.

In particular, as we will see, the explicit non-linear relations between different forms of the goldstino in the presence of gravity and matter are given by relations that express a given constrained goldstino superfield in terms of any other, among which a scalar nilpotent superfield is a direct superfield extension of the original Volkov-Akulov Lagrangian.
We will also show that the nilpotent chiral superfield studied in \cite{Casalbuoni:1988xh,Komargodski:2009rz} is reducible in the sense that it is the sum of the nilpotent chiral superfield of Ro\v{c}ek \cite{Rocek:1978nb} and another  chiral matter superfield satisfying a generalized nilpotency constraint.

The paper is organized according to its Table of Contents. We mainly use notation and conventions similar to that in \cite{Buchbinder:1995uq}.

\section{Brany nature of Volkov-Akulov model and its coupling to supergravity}\label{VA}
To construct the supersymmetric action for a spin-1/2 goldstone field Volkov and Akulov introduced \cite{Volkov:1972jx,Volkov:1973ix,Akulov:1974xz}, for the first time, the notion of superspace $M^{4|4}$ associated with the super-translation generators of the super-Poincar\'e algebra and parametrised by four bosonic space-time coordinates $x^m$ $(m=0,1,2,3)$ and four anti-commuting Weyl-spinor coordinates $\theta^\alpha$ and $\bar\theta^{\dot\alpha}$ ($\alpha,\dot\alpha=1,2$). The flat superspace coordinates transform under the Poincar\'e supersymmetry with parameters $\epsilon^\alpha$ and $\bar\epsilon^{\dot\alpha}$ in a conventional way
\bea\label{Psusy}
&\delta\theta^\alpha=\epsilon^\alpha,\qquad \delta\bar\theta^{\dot\alpha}=\bar\epsilon^{\dot\alpha},&\nonumber\\
&\delta x^{m}=\ii(\theta\sigma^m\bar\epsilon-\epsilon\sigma^m\bar\theta)\,.&
\eea
Next, Volkov and Akulov constructed the superinvariant Cartan one-form
\be\label{VAform}
E_0^m=\rd x^m+\ii(\theta\sigma^m \rd\bar\theta-\rd\theta\sigma^m\bar\theta)
\ee
and assumed that $\theta$ and $\bar\theta$ are actually fields in a four-dimensional subspace of $M^{4|4}$ depending on $x^m$.
In other words,
they considered a map of a four--dimensional surface $\mathcal M^4$ into the $M^{4|4}$ target superspace. A priori, the surface $\mathcal M^4$ can be parametrised by an independent set of four coordinates $\xi^i$ such that its embedding into  $M^{4|4}$ is described by the functions $x^m(\xi)$, $\theta^\alpha(\xi)$ and $\bar\theta^{\dot\alpha}(\xi)$. However, assuming the $\mathcal M^4$ diffeomorphism invariance of the embedding, one can always choose the $\mathcal M^4$ coordinate system (i.e. impose the so-called static gauge) in such a way that
\be\label{staticgauge}
x^m=\delta_i^m\,\xi^i.
\ee
In what follows we will use both the diffeomorphism invariant embedding description and the static gauge.

From the modern perspective this model describes just a space-filling 3-brane propagating in a flat $\mathcal N=1$, $D=4$ superspace and carrying the goldstone field
\be\label{chitheta}
 \chi^\alpha(x)=f\theta^\alpha(x),
 \ee
 where $f$ is a constant parameter of dimension of mass $m^2$ which characterizes the supersymmetry breaking scale (and the brane tension $T=f^2$).

Note that, at this level, such a brane can be considered as an auxiliary object which is useful for describing the goldstino and, as we will see, for coupling it to other supersymmetric or non-supersymmetric matter. On the other hand, this brane acquires a more fundamental  interpretation if the goldstino is associated to a physical brane in a higher-dimensional UV completion of the theory as has been extensively discussed in the literature (see e.g. \cite{McGuirk:2012sb,Bergshoeff:2013pia,Kallosh:2014wsa,Bergshoeff:2015jxa,Kallosh:2015nia,Aparicio:2015psl,Garcia-Etxebarria:2015lif,Dasgupta:2016prs,Vercnocke:2016fbt,Kallosh:2016aep} and references therein).

The field $\chi^\alpha(x)$ has the canonical dimension of $m^{\frac 32}$ and transforms under the supersymmetry variation \eqref{Psusy} non-linearly as a goldstone field
\be\label{susyv}
\delta\chi^\alpha=f\epsilon^\alpha-\delta x^m\partial_m\chi^\alpha=f\epsilon^\alpha
+\ii f^{-1}(\epsilon\sigma^m\bar\chi-\chi\sigma^m\bar\epsilon)\partial_m\chi^\alpha\,.
\ee
It is worth noting that the commutator of these transformations closes on space-time translations \emph{off the mass shell}, i.e. without the use of the goldstino equations of motion. This implies that there is no issue with the construction of supersymmetric couplings of the Volkov-Akulov goldstino to other fields which would otherwise require the use of auxiliary fields.

The supersymmetric Volkov-Akulov action is constructed as the 3-brane worldvolume integral
\be\label{VAa}
S_{\rm VA}=-f^2\int\d^4x \sqrt{-\det g_{mn}}=-f^2\int\d^4x \det  \mathbb E_{0\,m}{}^a,
\ee
where
\be\label{gmn}
g_{mn}=\mathbb E_{0\,m}{}^a\eta_{ab}\mathbb E_{0\,n}{}^b
\ee
is the induced worldvolume metric and
\be\label{E0mn}
\mathbb E_{0\,m}{}^a=\delta^a_m+\ii(\theta  \sigma^a\partial_m\bar\theta -\partial_m\theta \sigma^a\bar\theta)\, =\delta^a_m+\ii f^{-2}(\chi\sigma^a\partial_m\bar\chi-\partial_m\chi\sigma^a\bar\chi)\,
\ee
are the components of the pullback on the brane worldvolume of the Volkov-Akulov one-form \eqref{VAform}.

The leading terms in the action \eqref{VAa} are
\be\label{VAl}
S_{\rm VA}=-\int \d^4x (f^2+\ii\chi\sigma^m\partial_m\bar\chi-\ii\partial_m\chi\sigma^m\bar\chi+\ldots).
\ee
From the above expression we see that the overall sign in the Volkov-Akulov action is chosen in such a way that the goldstino kinetic term has the correct sign, then the first (constant) term in \eqref{VAl} becomes a positive (de Sitter) cosmological constant when the Volkov-Akulov action couples to supergravity. This explains the origin of the positive contribution to the cosmological constant in supergravity theories with spontaneously broken supersymmetry.

The action \eqref{VAa} is written in the static gauge \eqref{staticgauge}. Its worldvolume diffeomorphism invariant counterpart is
\be\label{VAdif}
S_{\rm VA}=-f^2\int\d^4\xi \det  E_{0\,i}{}^a(x(\xi),\theta(\xi),\bar\theta(\xi))\,.
\ee

Using the interpretation of the Volkov-Akulov action as that of the space-filling 3-brane, the {\em goldstino brane}, it is straightforward to couple it to $\mathcal N=1$, $D=4$ supergravity, for instance to the old minimal one, using the superfield approach \cite{Bandos:2015xnf}\footnote{In this paper we will restrict the consideration to the old minimal $\caln=1$ supergravity (except for a brief discussion of special minimal supergravity in Section \ref{gm}), but it can be straightforwardly extended to other off-shell supergravity multiplets by choosing appropriate sets of superfield supergravity constraints.}.  In the superfield formulation of supergravity the flat superspace vielbein \eqref{VAform} gets generalized to a curved superspace one
\be\label{Eaform}
E^a(z)=\rd z^{ M}E^a_{ M}(z)=\rd x^m E_m^a+\rd \theta^\mu E_\mu^a+\rd\bar\theta^{\dot\mu}E_{\dot\mu}^a, \qquad z^{ M}=(x^m,\theta^\mu,\bar\theta^{\dot\mu}),
\ee
where $a=0,1,2,3$ are vector tangent-space indices. The vector supervielbein $E^a(z)$ and its spinorial partners
\be\label{Ealpha}
E^\alpha(z)=\d z^{M}E^\alpha_{ M}(z), \qquad \bar E^\alpha(z)=\d z^{ M}\bar E^{\dot\alpha}_{M}(z)
\ee
are subject to certain torsion constraints. As components in their $(\theta,\bar\theta)$-expansion the supervielbeins contain the fields of the supergravity multiplet, the graviton $e^a_m(x)$, the gravitino $\psi_m^\alpha(x),\bar\psi_m^{\dot\alpha}(x)$,  the complex scalar auxiliary field $R(x)$ and the auxiliary vector field $G_a(x)$. The auxiliary fields are the leading components of the chiral scalar supergravity superfield $R(z)$  and the vector superfield $G_a(z)$, respectively \footnote{For details on the superfield description of $\mathcal N=1$, $D=4$ supergravity see e.g. \cite{Wess:1978bu,Ogievetsky:1978mt,Siegel:1978nn,Siegel:1978mj,Ogievetsky:1980qp,Howe:1981gz,Wess:1992cp} and \cite{Buchbinder:1995uq}. The earliest references on this subject are \cite{Akulov:1975ax,Akulov:1976ck,Ogievetsky:1976qc,Wess:1976fn,Wess:1977fn,Siegel:1977ng}.}. An explicit form of \eqref{Eaform} and \eqref{Ealpha} in a Wess-Zumino gauge  was computed to all orders in $\theta$s
and $\bar \q$s in \cite{Bandos:2015xnf}.

The coupling of the goldstino brane to supergravity is described by the following action \cite{Bandos:2015xnf}
\be\label{action1}
S=-\frac 3{\kappa^2} \int \d^8 z\, {\rm Ber}\,E - \frac m{\kappa^2} \Big(\int \d^6\zeta_L\, {\mathcal E}+\text{c.c.}\Big)-f^2\int \d^4\xi \det \mathbb E_i^a(z(\xi))
\; ,
\ee
where $\kappa^2$ is the gravitational coupling constant, the first term is the $\mathcal N=1$, $D=4$ supergravity action given as the volume of the full curved superspace $\mathcal M^{4|4}$ with ${\rm Ber}\,E$ being the superdeterminant of the supervielbein matrix $E^A_M$ \eqref{Eaform}-\eqref{Ealpha} \footnote{ ${\rm Ber}$ stands  for {\it Berezenian}, the name for the superdeterminant which is used to give credit to Felix Berezin, the founder of 'supermathematics'. },
the second term is a chiral superspace volume with a measure ${\mathcal E}$, $m$ is the gravitino mass which also defines the value of the supersymmetry preserving AdS cosmological constant $\lambda=-\frac{3m^2}{\kappa^2}$, and the third term describes the dynamics of the goldstino brane in curved superspace. Geometrically, the latter is the direct generalization of the flat space Volkov-Akulov action \eqref{VAdif} with $\mathbb E^a_i(z(\xi))$ being the pullback of the supervielbein \eqref{Eaform} on the 3-brane worldvolume, namely
\be\label{Eapull}
\mathbb E_i^a(z(\xi))=\partial_i x^m(\xi)E^a_m(z(\xi))+\partial_i\theta^\alpha(\xi) E^a_\alpha(z(\xi))
+\partial_i\bar\theta^{\dot\alpha}(\xi)E^a_{\dot\alpha}(z(\xi))\,.
\ee
The third term in \eqref{action1} is invariant under the worldvolume diffeomorphisms $\xi^{'i}=f^i(\xi)$  which, as we have already mentioned, can be used to identify the  worldvolume parameters $\xi^i$ with the space-time coordinates $x^m$ by imposing the static gauge \eqref{staticgauge}.

The local supersymmetry transformations of the goldstino field $\theta(x)=f^{-1}\chi(x)$ derived in \cite{Bandos:2015xnf} in the Wess-Zumino gauge have the following form
\be\label{varTh}
 \begin{aligned}
 \delta \theta^{\alpha}=& \epsilon^\alpha(x)   +\ii\left(\epsilon\sigma^m\bar\theta -\theta \sigma^m\bar\epsilon  \right)\; \left[\psi_m^\alpha +\nabla_m \theta{}^\alpha
 -\ii \left( \theta\sigma^n\bar\psi_m- \psi_m \sigma^n\bar\theta \right)(\psi_n^\alpha+\nabla_n\theta{}^\alpha)
 \right]\\ & -
 \frac 1{16} \left(\epsilon\sigma^a\bar\theta-\theta \sigma^a\bar\epsilon   \right)\; \left[2
 \theta^\alpha G_a + (\theta \sigma_{ab})^{\alpha} G^b+2 (\bar\theta\tilde\sigma_a)^{\alpha} R\right] +  \ldots\, , \qquad
\end{aligned}
\ee
where $\nabla_m=\partial_m+\omega_m(x)$ is a covariant derivative containing a spin connection $\omega^{ab}_m(x)$  and $\ldots$ stand for higher order terms in the fields.
Equation \eqref{varTh} reduces to \eqref{susyv} in the flat space limit.

\section{Description of the goldstino in terms of  constrained superfields}\label{CSF}

An alternative way to describe the goldstino is
to use
a superfield in which the only independent component is the goldstino itself while its superpartners are composites of the goldstino. So the goldstino superfield is constrained. A priori, the superfield transforms linearly under supersymmetry. The non-linear transformation of the goldstino is obtained by solving the superfield constraints. This construction is based on the general relation between linear and non-linear realizations of supersymmetry put forward in \cite{Ivanov:1977my,Ivanov:1978mx,Ivanov:1984hs} (see \cite{Ivanov:2016lha} for a recent review).

\subsection{Spinor goldstino superfields}

The goldstino may be embedded in a spinor superfield
as its lowest component in the expansion
in powers of the Grassmann variables.
In order for such a superfield to possess
no additional degrees of freedom,
its spinor covariant derivatives
must be some
functions of this superfield
and its spacetime derivatives.

A direct way of obtaining a constrained spinor superfield containing the goldstino is to act on the latter with a finite supersymmetry transformation whose parameters depend on the superspace coordinates \cite{Ivanov:1977my,Ivanov:1978mx,Samuel:1982uh}, {\it e.g.}
\be\label{spinorg}
\Xi_\alpha(x,\theta,\bar\theta)=\re^{\ri( \theta^\b Q_\b+\bar\theta_{\dot\b} \bar Q^{\dot \b})} \chi_\alpha(y)\,,\qquad x^m \equiv y^m+\ri f^{-2}\chi(y)\sigma^m\bar\chi(y)\, ,
\ee
where $y^m$ are complex coordinates, $Q^\alpha$ and $\bar Q_{\dot\alpha}$ are the supersymmetry generators, and $\x_\alpha(x)\equiv \chi_\alpha(y)$ is a ``chiral" goldstino whose supersymmetry variation involves only this field itself and not its complex conjugate
\be\label{chichiral}
\delta\x^\alpha=f\epsilon^\alpha-2\ri f^{-1}\x\sigma^m\bar\epsilon\,\partial_m\x_\alpha(x)\,.
\ee
This realisation for the goldstino was introduced by Zumino \cite{Zumino:1974qc},
and later it was exploited in \cite{Ivanov:1977my,Ivanov:1978mx,Rocek:1978nb,Samuel:1982uh}.
The superfield $\Xi_\a$  obeys the constraints \cite{Samuel:1982uh}
\bea \label{IKSW}
D_{\a} \Xi_{\b} &=&- f\ve_{\a \b } \, ,\qquad
\bar{D}_{\ad} \Xi_{\b} =  - {2 \ri}{f}^{-1} \Xi^{\a} \partial_{\a \ad} \Xi_{\b} \, .
\eea
This superfield was introduced for the first time by Ivanov and Kapustnikov
\cite{Ivanov:1978mx} although without technical details.
It was further elaborated by Samuel and Wess \cite{Samuel:1982uh}, including
its coupling to supergravity.

Alternatively, one can construct a spinor superfield directly from the original Volkov-Akulov goldstino  \cite{Ivanov:1978mx}  (see \cite{Wess:1992cp,Ivanov:2016lha} for reviews)
\be\label{Lam}
\Lambda_\alpha=\re^{\ri( \theta^\b Q_\b+\bar\theta_{\dot\b} \bar Q^{\dot \b})} \chi_\alpha(x)\,.
\ee
It obeys the constraints
 \bea
  \label{AV_Constraints}
	D_\a \L_\b =  - f\ve_{ \a \b} - \ri f^{-1} \bar\L^\ad \pa_{\a \ad}\L_\b\,,
	\qquad
	{\bar D}_\ad \L_\b = -\ri f^{-1} \L^\a \pa_{\a \ad} \L_\b \,.
\eea

One can also consider a chiral spinor goldstino superfield $\J_\a$  \cite{Ivanov:1978mx,Kuzenko:2011ya} (Ref. [23] added) subject to the constraints \cite{Kuzenko:2011ya} 
\bea\label{antichiral}
{D}_{\a} {\Psi}_{\b} =  -f \ve_{ \a \b }
+ 2 \ri f^{-1}\bar \Xi^{\ad} \partial_{\a \ad} {\Psi}_{\b}\, ,\qquad
\bar D_{\ad} {\Psi}_{\b} &=& 0 \,.
\eea
It may be shown that $\Xi_\a$ and $\Psi_\a$ are related to each other, and hence $D_\a \J_\b$ can be expressed solely in terms
of the superfields $\J_\g$ and $\bar \J_\bd$. However such an expression is less compact than the first relation in
\eqref{antichiral}.

The spinor fields $\x_\a(x) := \X_\a|_{\q=\bar \q =0}$ and
$\j_\a(x) := \J_\a|_{\q=\bar \q =0}$ naturally originate if one makes use of the coset parametrisation \cite{Kuzenko:2011ya}
\be
g\big(x, \xi (x), \bar{\psi} (x)\big)
= {\rm e}^{\ri( - x^a P_a + f^{-1}\xi^{\a}(x) Q_{\a})} \,
{\rm e}^{\ri f^{-1} \bar{\psi}_{\ad}(x) \bar{Q}^{\ad}}
\ee
in the framework of nonlinear realisations of
$\cN=1$ supersymmetry described in
\cite{Ivanov:1977my,Ivanov:1978mx,Samuel:1982uh}.
The fields $ \xi_{\a}$ and $ \bar{\psi}_{\ad}$ are related to the goldstino $\chi_{\a}$ and $ \bar{\chi}_{\ad}$ by
\be
\xi_{\a} (x) = \chi_{\a} (y)~, \qquad \bar{\psi}_{\ad}(x) = \bar{\chi}_{\ad} (y)~,\qquad y^m = x^m - \ri  f^{-2}\chi (y) \s^m \bar{\chi}(y).
\label{threeGoldstinos}
\ee
Conversely,
\be
\chi_{\a}(x) = \xi_\alpha(\hat y)~, \qquad \bar{\chi}_{\ad} (x) = \bar{\psi}_{\ad}(\hat y)~,\qquad \hat y^m = x^m + \ri f^{-2} \xi(\hat y) \s^m \bar{\psi} (\hat y).
\label{threeGoldstinos2}
\ee
These relations imply that the three different descriptions of the goldstino in terms of $\X_\a$, $\L_\a$ and $\J_\a$ are equivalent.
Given one of them, say $\X_\a$, the other superfields,
$\L_\a$ and $\J_\a$,
may be realised as composites of $\X_\a$, $\bar \X_\ad$ and their
covariant derivatives. For instance, the chiral spinor superfield $\J_\a$
is expressed in terms of $\X_\a$ and its conjugate
in a remarkably simple way:
\bea
\J_\a = -\frac{1}{4f^2}\bar D^2 (\X_\a \bar \X^2)\, .
\eea


\subsection{Scalar goldstino superfields}

There are three standard scalar superfields to describe
the goldstino. The oldest of them is the nilpotent chiral scalar $X$ introduced in \cite{Rocek:1978nb,Ivanov:1978mx}.
It obeys the constraints\footnote{The factor $-1/4$ in the last expression of \eqref{Rocek} is chosen for convenience, since in our conventions \hbox{ $-\frac 14{\bar D}^2\,\bar\theta^2=1$} and hence $-\frac 14  {\bar D}^2 \bar  X|_{\bar\theta=0}=\bar F_x$ singles out the auxiliary field component of $\bar X$.} \cite{Rocek:1978nb}
\bea \label{Rocek}
\bar D_{\dot\alpha} X =0\,,\qquad  X ^2=0\,,\qquad -\frac 14  X {\bar D}^2 \bar  X ={f} X \,.
\eea
Another option, which is naturally related to Ro\v{c}ek's
construction \cite{Rocek:1978nb}, is the
real scalar superfield $\cV$ introduced in \cite{Lindstrom:1979kq}. It is constructed
as the composite
\bea \label{3.14}
\cV = \frac{1}{f^2} \bar X  X\,
\eea
and obeys the constraints
\be\label{Vc}
 \mathcal V^2=0, \qquad
 \frac{1}{16}\mathcal VD^\alpha{\bar D}^2 D_\alpha \mathcal V
 = \mathcal V\,,
\ee
as well as some additional constraints which  will be discussed in more detail around eqs. \eqref{VDV}.
As will be shown in Section \ref{VA=sec}, the scalar superfield $\mathcal V$ is nothing but a superfield extension of the original Volkov-Akulov Lagrangian.

Equation \eqref{3.14} expresses $\cV$
as a descendant of $X$ and $\bar X$.
In its turn, $X$ can be thought of as a descendant of $\cV$, namely
\bea\label{XV}
 X = -\frac{f}{4}\bar D^2 \cV\, .
\eea

The third realisation is  a modified complex linear superfield introduced
in \cite{Kuzenko:2011ti}. It satisfies the following constraints
\be\label{cls}
-\frac 14 {\bar D}^2\Sigma={f}\,,\qquad
\S^2 =0\, , \qquad
-\frac 14 \Sigma{\bar D}^2D_\alpha\Sigma={f}
D_\alpha\Sigma\,.
\ee
The goldstino superfields $X$ and $\cV$ can both be read off from
$\S$ and $\bar \S$ as follows:
\bea \label{StoXV}
{f} X = - \frac{1}{4} \bar D^2 (\bar \S \S)\, ,
\qquad \cV = \frac{1}{f^2} \bar \S \S\, .
\eea
These relations show that, in a sense, $\S$ is the simplest
scalar goldstino superfield.

The first constraint in \eqref{cls} defines the so-called {\it modified}
complex linear superfield. In fact, the goldstino
can also be embedded in a standard complex linear superfield $\G$ ($\bar D^2 \G =0$), subject to additional constraints.
Such a goldstino superfield was constructed in 2011 by Tyler, as explained in \cite{Kuzenko:2015uca}.
 Later
 it was discussed, albeit in an incomplete form,
 in \cite{Farakos:2015vba}. The complete set of
the constraints is
 \bea
 \bar D^2 \G =0\,, \qquad \G^2 =0\,, \qquad
 -\frac 14  \G {\bar D}^2 \bar  \G ={f} \G \,,
\label{Tyler1}
\eea
where the last constraint was not given in \cite{Farakos:2015vba}.
This goldstino superfield  is naturally expressed in terms of $\S$ and its conjugate as follows \cite{Kuzenko:2015uca}:
\bea
\G = \bar \S - \frac{1}{4 f}  (\bar D_\ad \S ) \bar D^\ad \bar \S\,.
\label{Tyler2}
\eea


\subsection{Equivalence of the  goldstino superfields}\label{equivalence}

As one might already deduce from the discussion in the previous two subsections, all the spinor and scalar goldstino superfields considered therein
are equivalent to each other.
Given one of them,
e.g., the complex linear superfield $\S$ \eqref{cls}, the other goldstino superfields
may be realiased as composites constructed from
$\S$, its conjugate $\bar \S$ and their covariant derivatives. The equations \eqref{StoXV} provide
such relations for the goldstino superfield
$X$ and $\cV$.
We also can readily express the spinor goldstino superfield
$\X_\a$ defined by the constraints \eqref{IKSW}
in terms of $\S$ and $\bar \S$.
The corresponding relations were given in
\cite{Kuzenko:2011ti}. They are
\bea \label{StoX}
\X_\a = \frac{1}{{2}} D_\a \bar \S~, \qquad
\bar \X_\ad =\frac{1}{{2}} \bar D_\ad \S\, ,
\eea
On the other hand, the superfield $\S$ is constructed from
$\bar\X_\ad$ by the rule \cite{Kuzenko:2011ti}
\bea\label{S/X}
f \S = \bar \X_\ad \bar \X^\ad~.
\eea
It is worth comparing this simple result
with the expression for $X$ in terms of the chiral
spinor goldstino superfield $\J_\a$,
eq. \eqref{antichiral},
which was derived in \cite{Kuzenko:2011ya}:
\bea
f X =\J^\a \J_\a\,.
\eea

The above composites, which express one goldstino superfield
in terms of a different one, are polynomial.
A rational expression emerges if one wishes to express, e.g.,
$\X_\a$ via $X$. Making use of an observation in \cite{Cribiori:2016hdz}, we obtain
\bea\label{/XX}
 \X_\a = -2f \frac{D_\a  X}{D^2 X}\,.
\eea
The relations \eqref{S/X} and \eqref{/XX} allow us to express $\S$ in terms of $X$, $\bar X$ and their covariant derivatives, or in terms of $\mathcal V$ with the use of eq. \eqref{XV}, and so on and so forth.

The following comment is in order. As one could have noticed from the above relations, all the nilpotent scalar superfields are composites of spinor superfields. This just reflects a simple fact that in the physical theory in which (due to the spin statistics-correspondence) the spinor components form a basis of the odd elements of the Grassmann algebra, the even (e.g.) scalar nilpotent quantities should be composed of Grassmann-odd spinors in a Lorentz-covariant way. With this assumption, for instance the nilpotency constraint in \eqref{Vc} for the real scalar superfield $V$ is solved by the ansatz $\mathcal V=\digamma^\alpha \digamma_\alpha\bar \digamma_{\dot\a}\bar \digamma^{\dot\a}=C\bar C$, where $\digamma_\alpha$ is an arbitrary Grassmann-odd spinor superfield and $C=\digamma^2$ is a nilpotent complex scalar superfield of even Grassmann parity.

On the other hand, the constraint  $\mathcal V^2=0$ is also solved by $\mathcal V=\eta\bar\eta$, where $\eta$ is a  Grassmann-odd complex scalar superfield ($\eta^2\equiv 0$). To exclude such unphysical solutions from the consideration one should impose additional constraints on $\mathcal V$, which are identically satisfied by the physical solution $\mathcal V=C\bar C$ and are not satisfied by $V=\eta\bar\eta$. These are
\be\label{VDV}
\mathcal VD_A D_B\mathcal V=0\,,\qquad \mathcal VD_AD_BD_C\mathcal V=0\,,
\ee
where $D_A= (\pa_a, D_\a, \bar D_\ad) $.
Then, the constraints \eqref{Vc} accompanied by \eqref{VDV} single out $\mathcal V$ which is expressed in terms of the other goldstino superfields as discussed above.
One may check that the constraints \eqref{Vc} and \eqref{VDV} allow one to express all the components of $\cV$
in terms of the goldstino field  identified with
$-\frac{1}{4} \bar D^2 D_\a \cV|_{\q=0}$.

Making use of the constrains \eqref{Vc} and
\eqref{VDV}, one may show that
\bea\label{W4}
\cV = \cW^\a \cW_\a \bar \cW_\ad \bar \cW^\ad\,,
\qquad \cW_\a = -\frac{1}{4} \bar D^2 D_\a \cV\,.
\eea
The constraint, which one has to use in order  to prove this result, is
\bea
\cV \cW_\a =0\,,
\eea
which is a special case of \eqref{VDV}.

The relations between the different constrained superfields can be used to find in a straightforward way the non-linear field redefinitions from one realization of the goldstino to another. A general form of such field redefinitions was obtained in \cite{Kuzenko:2010ef,Kuzenko:2011tj} in a different way by comparing all the known component versions of the Volkov-Akulov action. Such a procedure can hardly be directly generalized to the case of couplings of the goldstini to supergravity and matter multiplets. On the other hand the use of the relations between the constrained superfields still allows one to get such relations. One should only properly generalize the constraints to the curved superspace \cite{Lindstrom:1979kq,Samuel:1982uh,Kuzenko:2011ti} as will be reviewed in Section \ref{gsugra}.

In spite of the fact that all the goldstino superfields are equivalent, some of them turn out to be preferable when one is interested, e.g., in their couplings to supergravity and supersymmetric matter. In this respect, the scalar goldstino superfields are more suitable than the spinor ones, as was already noticed in \cite{Samuel:1982uh}.
The goldstino superfields $X$ and $\cV$ were coupled  to pure supergravity in \cite{Lindstrom:1979kq}, and their
simple couplings
 to supersymmetric matter were given in \cite{Samuel:1982uh}.
The goldstino superfield $\S$ has been coupled
both to supergravity and chiral matter superfeilds \cite{Kuzenko:2011ti,Kuzenko:2015yxa}. We will consider these couplings in more detail in Sections \ref{gsugra} and \ref{sugra+Sigma}.

To conclude our review of the known goldstino superfields,
we give three different, but equivalent forms of the
goldstino action:
\begin{subequations}
\bea
S&=&-f^2\int \rd^4x \rd^2\theta \rd^2\bar\theta\, \mathcal V \label{3.24a}\\
&=& - \int \rd^4 x  \rd^2  {\theta} \rd^2 \bar \q\,
\bar \S \S  \label{3.24b}\\
&=& -\hf \int \rd^4 x  \rd^2  {\theta} \,{\Psi}^{\a} {\Psi}_{\a}
-\hf \int \rd^4 x  \rd^2  \bar{\theta} \,\bar{\Psi}_{\ad} \bar{\Psi}^{\ad} \,.
\label{3.24c}
\eea
\end{subequations}
The action \eqref{3.24a}, introduced in \cite{Lindstrom:1979kq}, has the form of the $\cN=1$
Fayet-Iliopoulos term. The action \eqref{3.24b}, introduced in \cite{Kuzenko:2011ti},
coincides with the kinetic term for a complex linear superfield. Finally, the action \eqref{3.24c},
introduced in \cite{Kuzenko:2011ya}, has the form of a mass term for the chiral
spinor superfield \cite{Siegel:1979ai}.

As a final comment, let us mention that a simple generalization of the spinor goldstino superfield $\X_\alpha$ to $\cN\geq 2$ spontaneously broken supersymmetry was carried out in \cite{Kandelakis:1986bj}. A more general class of constrained $\cN=2$ goldstino superfields was studied in \cite{Kuzenko:2011ya}. Other aspects of
$\cN\geq 2$ cases have been recently considered in \cite{Cribiori:2016hdz}.


\subsection{Constrained goldstino superfields in supergravity}\label{gsugra}

When coupling the goldstino superfields to  supergravity one should appropriately modify their rigid supersymmetry constraints and the relations between the different superfields  discussed in the previous Sections. The `general' rule, which works in most cases, is to replace flat superspace covariant derivatives  with their curved superspace counterparts and the chiral projector $\bar D^2$ with $\bar{\mathcal D}^2-4R$, namely
\be\label{DD}
\partial_a \to \mathcal D_a,\qquad D_\alpha \to \mathcal D_\alpha\,\qquad \bar D_{\dot\alpha}\to \bar{\mathcal D}_{\dot\alpha},\qquad \bar D^2 \to \bar{\mathcal D}^2-4R\,,
\ee
where $ (\mathcal D_a,\mathcal D_\alpha,\bar{\mathcal D}_{\dot\alpha})$ are supercovariant derivatives and $R(z)$ is the chiral scalar curvature superfield.
In this way Lindstr\"om and Ro\v{c}ek \cite{Lindstrom:1979kq} coupled to supergravity the superfields $X$ and $\mathcal V$:
\bea \label{X2g}
\bar \cD_{\dot\alpha} X =0\,,\qquad  X ^2=0\,,\qquad -\frac 14  X ({\bar \cD}^2 -4R)\bar  X ={f} X \,;
\eea
\be\label{Vccurved}
\mathcal V^2=0\,, \qquad
\frac{1}{16}\mathcal V\mathcal D^\alpha(\bar{\mathcal D}^2-4 R)\mathcal D_\alpha \mathcal V= \mathcal V\,.
\ee
Note that, like in flat superspace, the left hand side of the second expression in \eqref{Vccurved} is equal to its complex conjugate.

The relation \eqref{XV} between $X$ and $\mathcal V$ take the following form
\be\label{XV1}
X=-\frac f4(\bar{\mathcal D}^2-4R)\,\mathcal V\,.
\ee

The local supersymmetry modification of the constraints for the spinor goldstino superfield \eqref{IKSW} is as follows
\begin{subequations} \label{SW_SUGRA}
\bea
\bar \cD_\ad \bar \X_\bd &=& \ve_{\ad \bd}
\Big\{ f - f^{-1}R\, \bar \X^2\Big\}\,,  \label{SW1}\\
\cD_\a \bar \X_\bd &=& f^{-1}
\Big\{
2\ri \bar \X^\gd {\cD_a} \bar \X_\bd -\delta_\bd^\gd G_a \bar \X^2 \Big\}\sigma^a_{\a\gd}
\,.
\label{SW2}
\eea
\end{subequations}
These constraints\footnote{The $R$-dependent term in \eqref{SW1}
and the $G$-dependent term in \eqref{SW2}
are examples of the non-minimal contributions that cannot be obtained by making use of the minimal prescription \eqref{DD}.}
were obtained by
Samuel and Wess \cite{Samuel:1982uh}
as a result of a nontrivial guess work. A straightforward way to get them is to make use of the relation \eqref{StoX} between $\X_\alpha$ and the complex linear  superfield $\S$, and the constraints   \eqref{cls} satisfied by the latter, which in the supergravity case are modified as follows \cite{Kuzenko:2011ti}
\bea
\X_\a = \frac{1}{{2}} \cD_\a \bar \S~, \qquad
\bar \X_\ad =\frac{1}{{2}} \bar \cD_\ad \S \,,
\label{3.311}
\eea
with $\S$ obeying the constraints
\be\label{cls1}
-\frac 14 (\bar{\mathcal D}^2-4 R)\Sigma={f}\,,\qquad
\S^2=0\, , \qquad
-\frac 14 \Sigma(\bar{\mathcal D}^2-4 R)\mathcal D_\alpha\Sigma={f}\mathcal D_\alpha\Sigma\,.
\ee


\subsection{Reducible goldstino superfields}

All the goldstino superfields considered so far
are irreducible in the sense that they contain only one independent
component field -- the goldstino itself, while the remaining component fields are simply composites constructed from the goldstino.
There also exist reducible goldstino superfields that
contain several independent fields, one of which is the goldstino.
A given reducible goldstino superfield can always be represented
as an irreducible one plus a ``matter'' superfield, which contains
all the component fields except for the goldstino.

As an example of reducible goldstino superfields, here we
consider the nilpotent chiral scalar
$S$ studied in \cite{Casalbuoni:1988xh,Komargodski:2009rz}.
It only satisfies the constraint
 \be\label{S}
 \bar D_\ad S =0\, ,\qquad S^2=0\quad \Longrightarrow \quad
 S(x,\q, \bar \q)=\re^{\ri \q \s^a \bar \q \pa_a} \Big\{
 \frac{\l_s^2}{4F_s}+ \theta \l_s+ \theta^2 F_s \Big\}\, ,
 \ee
and thus differs from the superfield $ X $ in \eqref{Rocek}. In addition to the goldstino field\footnote{Upon elimination of the auxiliary field $F_s$,
the goldstino field $\l_s^\a$ is related to the Volkov-Akulov goldstino $\chi^\a$, eq. \eqref{chitheta}, by a nonlinear field redefinition \cite{Kuzenko:2010ef,Kuzenko:2011tj}.}
$\l_s{}^\a$ it has the independent auxiliary field $F_s(x)=-\frac 14 D^2S|_{\theta=\bar \q=0}$, which is required to be nowhere vanishing,
and hence $D^2 S \neq 0$. It follows from \eqref{S} that $S$ can be written
in the form
\bea
S=-\frac{(D^\a S) (D_\a S)}{D^2 S}\,.
\eea

It was  shown in \cite{Komargodski:2009rz,Kuzenko:2011tj} that, for the pure goldstino model,
the superfield $S$ coincides with $ X $ on the mass shell when $F_s(x)-f$ is expressed in terms of the goldstino and its derivatives. As we will see, this connection can be understood and generalised by expressing $S$ in terms of  $X$ and an additional auxiliary chiral superfield.

We start by showing, in the supergravity framework,  that
the nilpotent covariantly chiral scalar $S$,
\bea\label{cS}
\bar \cD_\ad S = 0\, , \qquad S^2=0
\eea
can be represented as a sum of two covariantly chiral superfields,
\be\label{X+Y}
S = X +Y
\ee
of which $X$ is the nilpotent goldstino superfield \eqref{X2g} and $Y$ is a chiral matter superfield
satisfying a generalized nilpotency constraint
\be\label{nullY}
  2 X Y + Y^2=0\,.
\ee
This condition extends the class of the nilpotent superfield constraints studied so far \cite{Brignole:1997pe,Komargodski:2009rz,Dall'Agata:2015lek,Dall'Agata:2016yof,Kallosh:2016hcm} and briefly discussed in Section \ref{sugra+VA}.

It can be directly checked that the spinor $\lambda_y$ and the auxiliary field $F_y$ of $Y$ which solves \eqref{nullY} are independent, while its scalar component $\phi_y$ is expressed in terms of the goldstino $\lambda_x$ of $X$, its derivatives and the fields $\lambda_y$ and $F_y$.

The arbitrariness of $\lambda_y$ in $Y$ can be fixed in terms of $F_y$ and $\lambda_x$ by expressing $X$ in terms of $S$ as follows. Let us first introduce the composite superfield
\bea\label{XiS}
\bar \X_\ad =-2f \frac{\bar \cD_\ad \bar S}{\bar \cD^2 \bar S}\,,
\eea
which reduces to the one constructed in \cite{Cribiori:2016hdz} in the flat superspace limit. It proves to obey the constraint \eqref{SW_SUGRA}.

We are now in a position to introduce  the chiral scalar $X$ as a function of \eqref{XiS}
by the standard rule
\be\label{XSigma}
f^3 X=- \frac{1}{4}(\bar \cD^2 -4R)
(\X^2 \bar \X^2)\,.
\ee
Thus, modulo supergravity fields, the independent goldstino $\lambda_x$ in $X$ is expressed in terms of the $\lambda_s$ goldstino, the auxiliary field $F_s$ of $S$ and their derivatives. The relation for $\lambda_x$ can be inverted in the sense that $\lambda_s$ can be expressed in terms of $\lambda_x$, $F_s$ and their derivatives. Then the components of the chiral superfield $Y$ in \eqref{X+Y} and \eqref{nullY} are univocally defined in terms of
$\lambda_x$ and $F_s=F_x+F_y$ by $Y=S-X$. As a result,  the only
independent component of $Y$ is the auxiliary field.

Another way to relate the superfields $S$ and $X$ is to notice that the nilpotency constraint in \eqref{S} is invariant under arbitrary rescaling of $S$ with an unconstrained chiral superfield parameter, let us call it again $Y$. Then we can always represent the superfield $S$ as follows
\be\label{YX}
S=YX.
\ee
Note that the superfield Y is determined modulo a gauge transformation
\be\label{YDY}
Y\quad \rightarrow \quad Y+\Delta Y,
\ee
where $\Delta Y$ is a chiral superfield satisfying the constraint $X\Delta Y=0$.

Consider now the action for the superfield $S$ \cite{Casalbuoni:1988xh,Komargodski:2009rz}
\be\label{Saction}
S=\int \rd^4x \rd^2\theta \rd^2\bar\theta \,S\bar S-\left(\int \rd^4x\rd^2 \theta\,f\,S\,+\,{\rm c.c.}\right).
\ee
We see that, assuming that the superfield $S$ be the composite \eqref{YX}, the action \eqref{Saction} describes the coupling of the nilpotent superfield $X$ to the auxiliary chiral superfield $Y$ with the K\"ahler potential $K=Y\bar Y X\bar X$ and the superpotential $W=-fYX$. The superfield $Y$ can be easily integrated out by solving its equation of motion
\be\label{Yeq}
-\frac 14 X{\bar D}^2(\bar Y\bar X)=fX\,.
\ee
Multiplying the above equation by $Y$ we see that $S=YX$ satisfies the same constraint as $X$, eq. \eqref{Rocek}. Therefore, as one can directly check, the general solution of \eqref{Yeq} is
\be\label{Ysol}
Y=1+C,
\ee
where the chiral superfield $C$ is constrained by the condition $CX=0$. Thus, on the mass shell $Y$ is equal to unity modulo the gauge transformation \eqref{YDY}.

Upon substituting this solution into \eqref{Saction}  the action
reduces to that for the irreducible nilpotent superfield $X$ \cite{Rocek:1978nb}. This demonstrates from yet another perspective the relation between the two descriptions of the goldstino.

In \cite{Casalbuoni:1988xh,Komargodski:2009rz} the nilpotent superfield $S$ was regarded to be the most suitable for the description of couplings of the goldstino to matter and supergravity multiplets, since if one deals with the other constrained superfields these couplings require modifications of their constraints, as was discussed in Section \ref{gsugra}. So, though the coupling of the constrained superfields $ X $
\eqref{Rocek}, $\mathcal V$ \eqref{Vc} and $\Xi_\alpha$ \eqref{spinorg} to supergravity was considered already in \cite{Lindstrom:1979kq,Samuel:1982uh,Ivanov:1984hs,Ivanov:1989bh} and their quite general couplings to matter in \cite{Samuel:1982uh}, the most of recent work on the description of spontaneous supersymmetry breaking in a generic matter-coupled $\mathcal N=1$, $D=4$ supergravity uses the nilpotent chiral superfield $S$ of \cite{Casalbuoni:1988xh,Komargodski:2009rz} (see e.g. \cite{Farakos:2013ih,Dudas:2015eha,Bergshoeff:2015tra,Hasegawa:2015bza,Ferrara:2015gta,Antoniadis:2015ala,Kallosh:2015tea,Kallosh:2015sea,Schillo:2015,Kallosh:2015pho,Ferrara:2016een} and references therein). A complication one should deal with in this case is the necessity to perform a non-Gaussian integration of the auxiliary field $F$ (or the superfield $Y$) when deriving the component actions that only involve physical fields (see \cite{Bergshoeff:2015tra,Hasegawa:2015bza,Schillo:2015,Kallosh:2015pho} and references therein).


\subsection{Goldstino brane Lagrangian and goldstino superfields}
\label{VA=sec}
To complete the discussion of the equivalence of the different formulations, we will now show that the nilpotent scalar superfield $\mathcal V$ \eqref{Vc} is nothing but a superfield extension of the original Volkov-Akulov Lagrangian \eqref{VAa} or its goldstino brane extension to supergravity \eqref{action1}.
To this end, let us remind that the different forms of the component goldstino action associated to one or another constrained superfield are obtained from the superfield action \eqref{3.24a}
where $\mathcal V$ is either considered as the constrained real scalar superfield \eqref{Vc}, or as the
composite constructed from
other constrained superfields, \emph{e.g.} \eqref{3.14} or \eqref{StoXV}.

When coupled to supergravity, the action \eqref{3.24a} takes the form
\be\label{Vactioncurved}
S=-f^2\int \rd^8z\,{\rm Ber}\,E\, \mathcal V\,.
\ee

Now we notice that also the original worldvolume diffeomorphism invariant Volkov-Akulov action  \eqref{VAdif} can be rewritten as an integral in the bulk superspace with the use of the Dirac delta-finction and the delta-functions of the Grassmann-odd coordinates $\delta^2(\theta-\theta_0)\equiv(\theta-\theta_0)^2$ and $\delta^2(\bar\theta-\bar\theta_0)\equiv(\bar\theta-\bar\theta_0)^2$
\be\label{VAa1}
S_{\rm VA}=-f^2\int \d^8 z\Big[ \int \rd^4\xi\,(\theta-\theta(\xi))^2(\bar\theta-\bar\theta(\xi))^2 \,\delta^4(x-x(\xi))\det  E_{0\,i}{}^a(z(\xi))\Big]\,.
\ee
Upon integrating the above expression over the worldvolume coordinates, which effectively picks up a static gauge, we get
\be\label{VAa2}
S_{\rm VA}=-f^2\int\rd^8z\,(\theta-f^{-1}\chi(x))^2(\bar\theta-f^{-1}\bar\chi(x))^2 \det  E_{0\,m}{}^a(\chi,\bar\chi)\equiv \int\rd^8z\,\mathcal V \,
\ee
in which we made the substitution
\be\label{tchi}
\theta(\xi(x))=f^{-1}\chi(x), \qquad \bar\theta(\xi(x))=f^{-1}\bar\chi(x).
\ee
In $\mathcal N=(1,1)$, $D=2$ case the Volkov-Akulov action was written in a  form similar to \eqref{VAa2} in \cite{Rocek:1978nb} and in the $D=4$ case the form of $\mathcal V$ defined in \eqref{VAa2} was discussed in \cite{Luo:2009ib}.

The generalization of \eqref{VAa1} and \eqref{VAa2} to the curved superspace goldstino brane action \eqref{action1} has the same form as \eqref{Vactioncurved}, where
\be\label{V=VAcov}
\mathcal V=\int \rd^4\xi \,\delta^8 (z-z(\xi))\,\frac{\det {\mathbb E}(z(\xi))}{{\rm Ber} E(z(\xi))}. \qquad
\ee
Finally, upon integrating \eqref{V=VAcov} over the worldvolume and using \eqref{tchi} we get
\be\label{V=VA}
\mathcal V=(\theta-f^{-1}\chi(x))^2(\bar\theta-f^{-1}\bar\chi(x))^2\,\frac{\det {\mathbb E}(z(x))}{{\rm Ber} E(z(x))}, \quad z(x)=(x,f^{-1}\chi(x),f^{-1}\bar\chi(x))\,.
\ee
By construction \eqref{V=VAcov} (or \eqref{V=VA}) transforms as a scalar superfield which can thus be identified with the constrained scalar superfield in \eqref{Vactioncurved}. Indeed, one can directly check (e.g. in the flat case $({\rm Ber}\,E=1$)) that the nilpotent superfield $\mathcal V$ taken in the form \eqref{V=VA} satisfies the constraints \eqref{Vccurved} and the relations to all the other constrained superfields discussed above. For instance, given \eqref{V=VA} one gets the chiral nilpotent superfield $ X=-\frac{1}{4}(\bar{\mathcal  D}^2-4R) \mathcal V$ satisfying \eqref{X2g}. So, working with the goldstino brane, or equivalently with $\mathcal V$ in the form \eqref{V=VA},   simplifies the construction since in this formulation all the superfield constraints are solved and one does not need to care about them  when considering a general coupling of the goldstino to supergravity in the presence of matter in the bulk (see \cite{Ferrara:2016een} for the discussion on the incorporation of the superfield constraints into the goldstino actions as Lagrange multiplier terms). In addition, as we will discuss in the following Sections, one can directly couple the goldstino to fields  propagating in the brane worldvolume such as Born-Infeld vector fields, and scalar and fermion modes associated with extra dimensions. This should be useful for establishing a more direct relation of brane constructions in string theory with four-dimensional effective field theory models of spontaneous supersymmetry breaking, which is a topic of a great interest and importance (see e.g. \cite{McGuirk:2012sb,Bergshoeff:2013pia,Ferrara:2014kva,Kallosh:2014wsa,Bergshoeff:2015jxa,Kallosh:2015nia,Aparicio:2015psl,Garcia-Etxebarria:2015lif,Dasgupta:2016prs,Vercnocke:2016fbt,Kallosh:2016aep}).

\section{Coupling the goldstino to old minimal supergravity with matter
}\label{sec:coupling}
We will now consider how the goldstino couples to old minimal $\mathcal N=1$, $D=4$ supergravity and matter supermultiplets in different descriptions of the former and show that these different descriptions result in similar supersymmetry breaking terms. In Subsection \ref{gm} we will also briefly discuss couplings of goldstino to a so called special minimal (three-form) supergravity \cite{Grisaru:1981xm,Ovrut:1997ur,Kuzenko:2005wh,Bandos:2011fw,Bandos:2012gz}.

Every off-shell $\cN=1$ supergravity-matter system may be reformulated as
a super-Weyl invariant theory at the cost of introducing
a superconformal compensator, which
is a nowhere vanishing covarantly chiral scalar superfield $Y$
in the case of old minimal supergravity.
This non-trivial statement has its origin from the prepotential formulation of minimal supergravity \cite{Siegel:1977hn,Siegel:1978nn,Ogievetsky:1978mt}
(see \cite{Buchbinder:1995uq} for a review).
The resulting action is invariant
under the following super-Weyl transformation
of the supervielbein $E_M^A$ and the compensator $Y$
\cite{Howe:1978km,Siegel:1978fc}:
\be\label{superWeyl}
\begin{aligned}
E_{ M}^a&\rightarrow  \re^{\Upsilon+\bar\Upsilon}\,E_{ M}^a\,,\\
E_{M}^\alpha& \rightarrow\  \re^{2\bar\Upsilon-\Upsilon}\,\big(E_{M}^\alpha-\frac{\ii}{2}E_{M}^a\sigma^{\alpha\dot\alpha}_a\bar{\mathcal D}_{\dot\alpha}\bar\Upsilon \big)\,,\\
Y&\rightarrow \re^{-2\Upsilon}Y.
\end{aligned}
\ee

At the classical level,
the 
practical virtue of
the super-Weyl invariant  reformulation of supergravity-matter systems
is that it allows one to couple supergravity to matter in a K\"ahler--invariant manner and
considerably simplifies
the reduction to component fields
by  imposing suitable super-Weyl gauge conditions on the components
of the chiral compensator $Y$.
The idea goes back to the work by Kugo and Uehara
\cite{Kugo:1982mr}, and its systematic use was made
in  \cite{Kuzenko:2005wh}
for four-dimensional  $\cN=1$ supergravity-matter systems
and also in \cite{Kuzenko:2013uya} for three-dimensional $\cN=2$ supergravity-matter theories.  The power of this approach is that
it allows one to automatically obtain canonically
normalised component actions in the Einstein frame
without going through a 
tedious procedure 
described, e.g., in \cite{Wess:1992cp}.
The super-Weyl invariant  reformulation of supergravity-matter systems is also useful at the quantum level, see, e.g.,
\cite{Kaplunovsky:1994fg}.

The super-Weyl- and K\"ahler-invariant superfield action describing the coupling of old minimal $\mathcal N=1$, $D=4$ supergravity to chiral matter $\Phi^I$ and non-Abelian gauge superfields $V=V^At_A$ (with $t_A$ being the generators of a gauge symmetry algebra) has the following form  \cite{Wess:1978bu,Wess:1992cp,Binetruy:2000zx}
\bea\label{mattercoupling}
S_{\rm bulk}&=&-\frac 3{\kappa^2} \int {\rm d}^8z \,{\rm Ber} \,E\,\re^{-\frac 13K(\bar\Phi,\Phi,V)}\,Y\bar Y\\
&&- \frac m{\kappa^2}\Big(\int {\rm d}^6\zeta_L \,{\mathcal E}\,Y^3\,\big[\,W(\Phi)+ g_{AB}(\Phi)\,\mathbb W^{A\alpha} \mathbb W^B_\alpha\big]+{\rm c.c.}\Big)\,,\nonumber
\eea
where $K(\Phi,\bar\Phi,V)$ is a gauge invariant extension of the K\"ahler potential $K(\Phi,\bar\Phi)$ of a manifold $\calm_{\rm chiral}$ parametrised by the lowest components of the chiral superfields\footnote{Globally, $\calm_{\rm chiral}$ is a Hodge-K\"ahler manifold. This means that   $\re^{K(\Phi,\bar\Phi)}$ can be identified with a well defined metric of a complex line bundle  $L_K$ on $\calm_{\rm chiral}$ with even first Chern class. Note that, in comparison with the Wess-Bagger book \cite{Wess:1992cp} we have absorbed he gravitational constant $\kappa^2$ to the definition of the  K\"ahler potential ($K=\kappa^2 K_{\rm WB}$).} $\Phi^I$,
$W(\Phi)$ is a gauge invariant superpotential, which is a  holomorphic section of the complex line bundle $L_K$ on $\calm_{\rm chiral}$. Finally,  $\mathbb W_\alpha\equiv \mathbb W^A_\alpha t_A=-\frac14(\bar\cald^2-4R)e^{-V}\cald_\alpha e^{V}$ is the gauge field strength superfield  and  $g_{AB}(\Phi)$ are locally holomorphic functions which transform appropriately under the action of the gauge group and whose transition functions along $\calm_{\rm chiral}$ may involve non-trivial dualities of the vector multiplets.

The action  is invariant under   K\"ahler transformations and super-Weyl transformations. A general K\"ahler transformation acts on the K\"ahler potential, the superpotential and the conformal compensator as follows (for simplicity in what follows we skip the dependence of $K$ on $V$)
\be\label{supKahler}
\begin{aligned}
K(\Phi,\bar\Phi)&\rightarrow K(\Phi,\bar\Phi)+F(\Phi)+\overline{F(\Phi)}\,,\\
W(\Phi)&\rightarrow \re^{- F(\Phi)}W(\Phi)\,,\\
Y&\rightarrow \re^{\frac{1}3 F(\Phi)}Y,
\end{aligned}
\ee
where $F(\Phi)$ is a holomorphic function of $\Phi$.
\if{
On the other hand a general super-Weyl transformation, parametrised by a chiral  superfield $\Upsilon$ ($\bar\cald_{\dot\alpha}\Upsilon=0$), acts on the supervielbein and the conformal compensator as follows  \cite{Howe:1978km}
\be\label{superWeyl}
\begin{aligned}
E_{ M}{}^a&\rightarrow  e^{\Upsilon+\bar\Upsilon}\,E_{\mathcal M}{}^a\,,\\
E_{M}{}^\alpha& \rightarrow\  e^{2\bar\Upsilon-\Upsilon}\,\big(E_{M}{}^\alpha-\frac{\ii}{2}E_{M}{}^a\sigma^{\alpha\dot\alpha}_a\bar{\mathcal D}_{\dot\alpha}\bar\Upsilon \big)\,,\\
Y&\rightarrow e^{-2\Upsilon}Y.
\end{aligned}
\ee
This transformation does not effect the chiral superfields $\Phi^I$ and the vector superfields $V^A$.
}
\fi
The matter chiral superfields $\Phi^I$ and the gauge superfields $V^A$ are inert under the super-Weyl transformations.

The conformal compensator $Y$ is a pure gauge field. By gauge fixing the Weyl symmetry it can be reduced e.g. to
\be\label{Y1gauge}
Y=1\,.
\ee
This gauge is invariant under the residual combined K\"ahler-Weyl transformation with
\be\label{FSigma}\nonumber
F=6 \Upsilon.
\ee
However, in general the gauge (\ref{Y1gauge}) does not lead to a component supergravity action that is canonically normalised in the standard Einstein frame. The latter is recovered by  performing  a further Weyl re-scaling of the vielbein
 \be\label{We}\nonumber
 e^a_m(x)\,\rightarrow \,e^a_m(x)\, e^{\frac{1}6 K(\Phi,\bar\Phi,V)|_{\vartheta=0}}
 \ee
accompanied by an appropriate re-scaling of the fermionic fields (see e.g. \cite{Wess:1992cp}).

More practically, one can directly obtain the  Einstein frame description by fixing the gauge  \cite{Kugo:1982mr,Kaplunovsky:1994fg}
\be\label{EF}
\log Y +\log \bar Y=\frac{1}3 K(\Phi,\bar\Phi)|_{\rm harm}
\ee
where $|_{\rm harm}$ selects  the components of $K(\Phi,\bar\Phi)$  that can be written as the sum of a chiral and anti-chiral superfield. This is similar to the Wess-Zumino gauge in the case of the Abelian gauge pre-potential superfield $V$ whose lowest components are removed by an appropriate gauge transformation $V'=V-\Lambda-\bar\Lambda$.

\subsection{Reducible nilpotent chiral superfield couplings}\label{sugra+S}

A general coupling of the nilpotent chiral goldstino superfield $S$ \eqref{S} to old minimal supergravity and unconstrained matter and gauge multiplets is described by the action \cite{Hasegawa:2015bza,Kallosh:2015tea,Kallosh:2015sea,Schillo:2015} which we present in the superfield form\footnote{In \cite{Hasegawa:2015bza,Kallosh:2015tea,Kallosh:2015sea,Schillo:2015} the component counterpart of this action was constructed using the superconformal tensor calculus reviewed in \cite{Freedman:2012zz}.}
\bea\label{Smattercoupling}
S_{{\rm bulk}+S}&=&-\frac 3{\kappa^2} \int {\rm d}^8z \,{\rm Ber} \,E\,e^{-\frac{1}3\hat K(\bar\Phi,\Phi,S,\bar S)}\,Y\bar Y\\
&&- \frac m{\kappa^2}\Big(\int {\rm d}^6\zeta_L \,{\mathcal E}\,Y^3\,\big[\,\hat W(\Phi,S)+\,  \hat g_{AB}(\Phi ,S)\,\mathbb W^{A\alpha} \mathbb W^B_\alpha+ \Lambda S^2\big]+{\rm c.c.}\Big),\nonumber
\eea
where $\Lambda$ is the Lagrange multiplier taking care of the constraint $S^2=0$, and $\hat K$, $\hat W$ and $\hat g$ includes $S$ (and $\bar S$). Due to the nilpotency constraint, they have the following most general form \cite{Schillo:2015}
\be\label{hatK}
\hat K(\Phi,\bar\Phi, S,\bar S)=K(\Phi,\bar\Phi)+K_s(\Phi,\bar\Phi)S+\bar K_{\bar s}(\Phi,\bar\Phi)\bar S+  K_{s\bar s} (\Phi,\bar\Phi)S\bar S\, ,
\ee
\be\label{hatW}
\hat W(\Phi, S)=W(\Phi)+W_s(\Phi)S\, ,
\ee
\be\label{hatg}
\hat g_{AB}(\Phi, S)=g_{AB}(\Phi)+g_{AB}^s(\Phi)S\, .
\ee
It is instructive to see, following \cite{Komargodski:2009rz} in the flat case, how the equations of motion of the superfield $S$ produce modified constraints of the nilpotent superfield $X$ of \cite{Rocek:1978nb,Lindstrom:1979kq}.
The variation of \eqref{Smattercoupling} with respect to the Lagrange multiplier produces  the nilpotency condition $S^2=0$ and the variation with respect to $S$ gives the equation of motion
\bea\label{Sequation}
-\frac 14 (\bar{\mathcal D}\bar{\mathcal D}-4R){\Big(}e^{-\frac{1}3K}(K_{s\bar s}-\frac {1}{3} K_s\bar K_{\bar s})\bar SY\bar Y{\Big)}=\frac 14 (\bar{\mathcal D}\bar{\mathcal D}-4R)\Big(e^{-\frac{1}3K}K_sY\bar Y\Big)\nonumber\\
+{m}(W_s+g^s_{AB}\mathbb W^{A} \mathbb W^B+2\Lambda S)Y^3\,.\,\,\,\,\,\,\,
\eea
Multiplying the both sides of \eqref{Sequation} by $S$ and taking into account its nilpotency we get
\bea\label{Sequation1}
&-\frac 14 S(\bar{\mathcal D}\bar{\mathcal D}-4R){\Big(}e^{-\frac{1}3K}(K_{s\bar s}-\frac {1}{3} K_s\bar K_{\bar s})\bar SY\bar Y{\Big)}& \nonumber\\
&=S\Big[\frac 14 (\bar{\mathcal D}\bar{\mathcal D}-4R)\Big(e^{-\frac{1}3K}K_sY\bar Y\Big)+ {m}(W_s+g^s_{AB}\mathbb W^{A} \mathbb W^B)Y^3\Big]\,.\,\,\,\,&
\eea
Comparing eq. \eqref{Sequation1} with the third constraint in \eqref{X2g} we see that the former can be regarded as a modification of the Ro\v{c}ek superfield constraint in the presence of the matter fields thus providing the relation between the two realizations of the goldstino in which the auxiliary  field $F_S$ is on the mass shell.

The component structure of the action \eqref{Smattercoupling}, especially its non-linear dependence on the goldstino field $\chi(x)$ is very complicated \cite{Schillo:2015}. However the contribution into the action  of the goldstino sector drastically simplifies in the unitary gauge in which, using the local supersymmetry transformation, one sets the goldstino to zero\footnote{Note that, in general,  $\chi(x)$ is not exactly the physical goldstino, since the latter is a combination $\hat\chi$ of $\chi$ with other spinorial fields in the theory defined by the form of the gravitino mixing term $\psi_m\gamma^m \hat\chi$ in the action (\ref{mattercoupling}).}${}^,$\footnote{In Appendix A we will remind the well known fact that it is consistent to use the unitary gauge and set  the goldstone fields to zero directly in the locally symmetric actions. This does not result in the loss of the goldstone field equations, since they are not independent, but are consequences of physical field equations. This can also be seen at the path-integral level, since  from (\ref{varTh}) it follows that the Faddeev-Popov measure associated with (\ref{chi=0}) is a constant determinant and hence does not require the introduction of ghosts. }
\be\label{chi=0}
\chi^\alpha(x)=0=\bar\chi^{\dot\alpha}(x).
\ee
In this gauge the nilpotent superfield \eqref{S} reduces to $S=\theta^2 F_s$. So one can easily integrate the part of the action \eqref{Smattercoupling} containing $S$ over $\theta$, $\bar\theta$, $F_s$ and $\bar F_{\bar s}$ getting
\bea\label{S+US}
& S_{{\rm bulk}+S}=-\frac 3{\kappa^2} \int {\rm d}^8z \,{\rm Ber} \,E\,e^{-\frac{1}3K(\bar\Phi,\Phi,V)}\,Y\bar Y &\\
&- \frac m{\kappa^2}\Big(\int {\rm d}^6\zeta_L \,{\mathcal E}\,Y^3\,\big[\,W(\Phi)+ g_{AB}(\Phi)\,\mathbb W^{A\alpha} \mathbb W^B_\alpha\big]+{\rm c.c.}\Big)-\int {\rm d}^4x\,(\det e)\, U_S\,,\nonumber &
\eea
where $\det e\equiv \det e^a_m(x)$ and $U_S$ is the supersymmetry breaking potential produced by the nilpotent goldstino superfield couplings which (upon gauge fixing the conformal compensator as in \eqref{EF}) has the following form
\be\label{gV}
U_S=  \frac{|\frac 14(\bar{\mathcal D}\bar {\mathcal D}-4R)K_s +{m}e^{\frac {K}2 }\,(W_s+g^s_{AB}\mathbb W^{A\alpha}\mathbb W_\alpha^B)|^2}{ K_{s\bar s}-\frac {1}3 K_s\bar K_{\bar s} }|_{\theta=\bar\theta=0}\,.
\ee
We see that since $K_{s\bar s}$ is an arbitrary real and $K_s$ is an arbitrary complex gauge invariant function of $\Phi$ and $\bar \Phi$,  and  $W_s$ and $g^s_{AB}$ are arbitrary complex holomorphic functions of the matter fields
(more precisely one should speak of a section of complex line bundle, see paragraph below eq. (\ref{mattercoupling})), the potential \eqref{gV} gives rise to a very wide class of supergravity models exhibiting spontaneous supersymmetry breaking.

The action \eqref{Smattercoupling} can be further generalized by considering some of $\Phi$ be constrained superfields themselves \cite{Ivanov:1978mx,Brignole:1997pe,Komargodski:2009rz}. When searching for phenomenologically relevant effective field theory models the use of additional constrained superfields amounts to an effective removal from the consideration of extra heavy mass modes in an infinite mass limit (see e.g. \cite{Brignole:1997pe,Komargodski:2009rz,Dall'Agata:2015lek,Dall'Agata:2016yof,Kallosh:2016hcm} for more details and references). A ``master" constraint, proposed in \cite{Dall'Agata:2016yof}, which generates all known examples of the constrained superfields, except for \eqref{nullY},  has the following form
\be\label{SSQ}
\bar SS\,Q=0,\qquad S^2=0\,,
\ee
where $Q$ is a generic complex superfield which may carry external Lorentz indices.

A characteristic example is a  chiral superfield $T=\phi+\ii b+\theta\lambda +\theta^2 F_T$ satisfying the constraint \cite{Komargodski:2009rz}
\be\label{ST}
\frac 1{2\ii}S(T-\bar T)=0,\qquad S^2=0.
\ee
The superfield $T$ appears in interesting inflationary models \cite{Ferrara:2015tyn,Carrasco:2015iij,Kallosh:2016hcm} as an inflaton  supermultiplet in which sinflaton $b(x)$, sinflatino $\lambda(x)$ and the auxiliary field $F_T(x)$ are composites of the goldstino multiplet and the inflaton $\phi(x)$. In the unitary gauge $\chi(x)=0$ all the components of $T$ vanish except for the inflaton.


\subsection{Couplings of the irreducible constrained superfields}\label{sugra+Sigma}


The constrained superfield approach to spontaneously
broken $\cN=1$ supergravity was introduced by Lindstr\"om and Ro\v{c}ek in 1979 \cite{Lindstrom:1979kq}.
They considered the action
\bea
S = -
\int
\rd^8z
\,
{\rm Ber\,} E\left( \frac{3}{ \k^2} + \bar X  X \right)
- \left\{   \frac{ m}{ \k^2}
\int
\rd^6 \z_L
\,
{\cal E} + {\rm c.c.} \right\}
~, \label{4.19}
\eea
which describes the coupling of the old minimal $\cN=1$ supergravity to the goldstino superfield $X$ constrained
according to \eqref{X2g}.
They also used the equivalent form for the action,
which is obtained by replacing $\bar X X$ with
$f^2 \cV$, with $\cV$ constrained according to  \eqref{Vccurved}.
In the unitary gauge,
the cosmological constant was shown to be equal to
\bea
\Lambda =f^2 - 3\frac{ m^2}{{ \k}^2}~,
\label{cosmcon}
\eea
which coincides with the value obtained by Deser and Zumino
\cite{Deser:1977uq} in their study of the coupling of the pure
$\cN=1$ supergravity without auxiliary fields to the Volkov-Akulov action.
As was already mentioned, the positive contribution $f^2$ to the cosmological constant \eqref{cosmcon}, which comes from the goldstino superfield $X$, is universal for all superfield models of spontaneously broken $\cN=1$ supergravity
including those advocated in \cite{Bergshoeff:2015tra,Hasegawa:2015bza,Kuzenko:2015yxa,Bandos:2015xnf}.

The super-Weyl invariant reformulation of
\eqref{4.19} is described by the action
\bea
S = -
\int {\rm d}^{8} z\,
{\rm Ber\,} E\, \bar Y Y
\left( \frac{3}{ \k^2}  + \bar X  X \right)
-\left\{   \frac{m}{ \k^2}
\int {\rm d}^{6} \z_L\,
{\cal E} \,Y^3 + {\rm c.c.} \right\}
~, ~~~
\label{4.21}
\eea
where $X$ is chosen to be inert under the super-Weyl transformation.
The goldstino superfield $X$ now obeys the super-Weyl invariant constraints
\bea
X^2 =0~, \qquad
-\frac{1}{4} X (\bar \cD^2 -4R) (\bar X \bar Y Y)
= f Y^3 X ~.
\label{4.22}
\eea

Lindstr\"om and Ro\v{c}ek \cite{Lindstrom:1979kq}
did not discuss matter couplings for the goldstino superfield. To introduce such couplings,
it suffices to deform the constraints \eqref{4.22}
as follows:
\bea
X^2 =0~, \qquad
-\frac{1}{4} X (\bar \cD^2 -4R) (\bar X \mathcal F\bar Y Y)
= \mathcal W \,Y^3  X ~, \qquad \bar \cD_\ad \mathcal W =0\, .
\label{4.23}
\eea
Here $\mathcal F$ and $\mathcal W$ are {\it composite} real and covariantly chiral scalars, respectively, which are constructed from matter superfields. They both are chosen to be super-Weyl inert. The goldstino-dependent part of the action \eqref{4.21} should also be deformed,
\bea
\int {\rm d}^{8} z\,
{\rm Ber\,} E\, \bar Y Y  \bar X  X \quad
\Longrightarrow \quad
\int {\rm d}^{8} z\,
{\rm Ber}\,E\, \bar Y Y  \mathcal F \bar X  X \,.
\eea

Note that the constraint \eqref{Sequation1} is of the type \eqref{4.23} with specially chosen
composites $\mathcal F$
and $\mathcal W$. This means that the two formulations in terms of the different goldstino superfields $S$ and $X$ are equivalent.

To obtain a super-Weyl invariant formulation
for the complex linear goldstino superfield
coupled to supergravity, the first and third constraints in \eqref{cls1}
have to be deformed as follows
\cite{Kuzenko:2015yxa}
\bea
-\frac 14 (\bar{\mathcal D}^2-4 R)\Sigma={f}Y^2\,,
\qquad
-\frac 14 \Sigma(\bar{\mathcal D}^2-4 R)\mathcal D_\alpha \left(\frac{\Sigma}
{\bar Y}\right)
={f}Y^2\mathcal D_\alpha
\left(\frac{\Sigma}{\bar Y}\right)\,,
\label{4.25}
\eea
while keeping the condition $\S^2=0$ intact.
The super-Weyl transformation of $\S$ is chosen\footnote{Applying a field redefinition
$\S \to Y^n \S$ leads to a different super-Weyl transformation law of $\S$
and modifies the explicit form of constraints  \eqref{4.25}.}
to be
\bea
\S~ \to ~\re^{-2\bar \U} \S\,.
\eea
The action for supergravity
coupled to
this goldstino superfield
is obtained
from \eqref{4.21} by replacing
$\bar Y Y \bar X X ~\to ~ \bar \S \S $,
\bea
S = -
\int {\rm d}^{8} z\,
{\rm Ber\,} E\, \left\{
 \frac{3}{ \k^2} \bar Y Y + \bar \S \S\right\}
-\left\{   \frac{m}{ \k^2}
\int {\rm d}^{6} \z_L\,
{\cal E} \,Y^3 + {\rm c.c.} \right\}
~.
\label{4.255}
\eea
This action, whose form is an obvious corollary of the analysis carried out in \cite{Kuzenko:2011ti}, was explicitly given  in \cite{Kuzenko:2015yxa}. It
describes the minimal coupling of the complex linear goldstino superfield to supergravity
(compare with the action \eqref{3.24b} describing
the dynamics of the complex linear goldstino superfield in Minkowski superspace).

The complex linear goldstino superfield can
also  be coupled to supersymmetric matter
\cite{Kuzenko:2011ti}. The corresponding constraints
are obtained by deforming \eqref{4.25} to the form
\begin{subequations} \label{4.266}
\bea
&&-\frac 14 (\bar{\mathcal D}^2-4 R)\Sigma=\cW Y^2\,,
\label{4.266a}\\ \S^2 =0~, \qquad
&&
-\frac 14 \Sigma(\bar{\mathcal D}^2-4 R)\mathcal D_\alpha \left(\frac{\Sigma}
{\bar Y}\right)
={\cW}Y^2\mathcal D_\alpha
\left(\frac{\Sigma}{\bar Y}\right)\,.~~~
\label{4.266b}
\eea
\end{subequations}
where $\cW$ is a {\it composite} covariantly chiral scalar
chosen to be inert under the super-Weyl transformations.\footnote{As
discussed in \cite{Tyler:2014pwa}, the most general expression for the composite $\cW$
is as follows: $\cW = f + G_1 (\Phi) + G_2(\Phi)
{\rm tr}
({\mathbb W}^\a {\mathbb W}_\a)$,
where the matter chiral superfields $\Phi$ and ${\mathbb W}_\a$
were introduced at the beginning of this section.}
As the goldstino-dependent part of the supergravity-matter action, we may choose
\bea
S_\S = \int \rd^8 z\,
{\rm Ber\,} E\, \Big\{
- \bar \S {\mathcal F}^{-1} \S +Y \S \,\cC + \bar Y \bar \S \,\bar \cC \Big\}~,
\label{4.277}
\eea
where $\mathcal F$ and $\mathcal C$ are real and complex composite scalars,
respectively, which are chosen to be inert under the super-Weyl transformations.

If \eqref{4.266a} is the only constraint imposed on (non-nilpotent) $\S$,
the theory with action \eqref{4.277} possesses a dual formulation given by
\bea
S_{\cS} = \int \rd^8 z \,
{\rm Ber\,}E\,\bar Y Y
( \bar \cS +\bar \cC)  {\mathcal F} (\cS+  \cC)
-\left\{
\int
\rd^6 \z_L
\,
{\cal E} \,Y^3 \cS \cW + {\rm c.c.} \right\}~.~~~
\label{4.288}
\eea
Here the dynamical variable $\cS$ is a covariantly chiral scalar, $\bar \cD_\ad \cS =0$, chosen to be inert under  the super-Weyl transformations.
This action with $\cS$ constrained to be nilpotent, $\cS^2=0$,
is analogous to the part of \eqref{Smattercoupling}
containing the goldstino superfield $S$
subject to the constraint $S^2=0$.
The duality between \eqref{4.277} and \eqref{4.288}
indicates that, upon imposing the constraints \eqref{4.266b},
the goldstino-matter coupling described by \eqref{4.277}
is analogous to the one generated by the chiral action
\eqref{Smattercoupling}.

It is of interest to look at the properties of the spinor
goldstino superfield $\bar \X_\ad $ defined by \eqref{3.311}
in the case of the deformed constraints \eqref{4.266}.
To simplify explicit expressions, we will temporarily
choose the super-Weyl gauge $Y=1$. Making use of the condition
$\S^2=0$ gives
\bea
\S = \frac{1}{\cW} \bar \X_\ad \bar \X^\ad\,.
\eea
From \eqref{4.266a} we deduce
\bea
\bar \cD_\ad \bar \X_\bd &=& \ve_{\ad \bd}
\Big\{ \cW - \frac{R}{\cW}\, \bar \X^2\Big\}\,,
\eea
which is a generalisation of \eqref{SW1}. Making use of
\eqref{4.266b} one may work out a generalisation of \eqref{SW2}.
In the flat superspace limit, it is given by eq. (6.3.125)
in \cite{Tyler:2014pwa}.

There exist more general deformations of the constraints
on $\S$ than those given by \eqref{4.266}. We will not pursue
this topic in the present paper.

Finally, in the case of the real scalar superfield $\mathcal V$ of the Weyl weight 2, {\it i.e.}
$$\mathcal V \quad \to \quad \re^{2(\Upsilon+\bar\Upsilon)}\mathcal V$$
which, as we have seen, is related to the Volkov-Akulov Lagrangian in the most direct way, the coupling of the goldstino to supergravity with matter is described by the action
\be\label{Vacgm}
S_{\rm VA}=-f^2\int {\rm d}^8z\, {\rm Ber}\,E \,{\mathcal F}\,\bar Y^2 Y^2\,\mathcal V,
\ee
where $\mathcal V$ satisfies the constraints
\bea\label{Vccurved=2W}
&\cV^2=0\,, & \\
&\cV \cD_A \cD_B \cV =0~,
\qquad \cV \cD_A \cD_B \cD_C \cV=0\,,
\label{4.333}&\\ \label{4.32*}
&\frac{1}{16}{\mathcal V}{\mathcal D}^\alpha\bar{\mathcal D}^2\,\mathcal  D_\alpha \mathcal  V =\mathcal V\,.&
\eea
As we discussed in Section \ref{equivalence}, the constraints \eqref{4.333} generalizing eqs. \eqref{VDV}  single out the nilpotent $\mathcal V$ of the form \eqref{W4} which is directly related to the goldstino brane construction (see \eqref{V=VA}). The constraints \eqref{4.333} are super-Weyl invariant due to $\cV^2=0$ and imply that also \eqref{4.32*} is super-Weyl invariant. So the coupling of this goldstino superfield to matter does not require the modification of its constraints by including matter superfields, in contrast to the other cases of the irreducible goldstino superfields considered above.
\footnote{Note that for the nilpotent superfield which does not obey \eqref{4.333} the Weyl and K\"ahler invariant counterpart  of the constraint \eqref{4.32*} is $\frac{1}{16}{\mathcal V}{\mathcal D}^\alpha(\bar{\mathcal D}^2-4 R)\,\mathcal  D_\alpha (\cF\bar{Y}Y\mathcal  V )= \cF\,\bar{Y}Y \mathcal V$. }


\subsection{Goldstino brane couplings}\label{sugra+VA}
We will now show in detail that in the geometric framework of the non-linear realizations, one gets a similar general class of models of spontaneously broken matter-coupled supergravity as the ones discussed in the previous subsections, as well as new couplings which have not been discussed in the literature so far.

A general coupling of the goldstino brane to supergravity with matter proposed in \cite{Bandos:2015xnf} is described by the following action
\be\label{VAmattercoupling}
\begin{aligned}
S_{\rm bulk+brane}&=S_{\rm bulk}+S_{\rm brane}\\
&\equiv
S_{\rm bulk}-{f^2} \int {\rm d}^4\xi \,\det\big[\mathbb{E}(z(\xi))\big]\, \hat{\mathcal F}_{\Phi,\bar\Phi,V,\varphi }(\xi)
\; ,
 \end{aligned}
\ee
where $S_{\rm bulk}$ is as in (\ref{mattercoupling}), the goldstino appears only in $S_{\rm brane}$  (as in eqs. \eqref{action1}-\eqref{Eapull}) and
\be\label{hatF}
 \hat{\mathcal F}_{\Phi,\bar\Phi,V,\varphi}(\xi)\equiv\hat{\mathcal F}[\Phi(z(\xi)),\bar\Phi(z(\xi)),V(z(\xi)),\varphi(\xi)]
\ee
is a real supersymmetric, gauge and worldvolume diffeomorphism invariant function of the pull-backs of bulk superfields and their derivatives, as well as of purely brane worldvolume fields $\varphi(\xi)$ such as scalar, spinor or Born-Infeld vector fields, which may be regarded as those of a dimensionally reduced (anti)-D-brane of type II string theory as we will discuss in mored detail in Sections \ref{wv} and \ref{aD3}.

The goldstino brane term in \eqref{VAmattercoupling} can be extended to the integral over the curved superspace as in \eqref{Vactioncurved}, \eqref{V=VAcov}, \eqref{V=VA} or \eqref{Vacgm}, but this is not necessary, and actually redundant, for the analysis that follows.

The invariance of  (\ref{mattercoupling}) under the transformations \eqref{supKahler}-\eqref{superWeyl} ensures that \linebreak
$S_{\rm bulk}$ is globally well defined along the target space $\calm_{\rm chiral}$ of the chiral fields. This  property should also be satisfied by the brane action $S_{\rm brane}$ in  (\ref{VAmattercoupling}). Namely, since under the super-Weyl transformations \eqref{superWeyl} the brane density $\det\big[\mathbb{E}(z(\xi))\big]$ has weight 4, the function \eqref{hatF} should contain the compensating factor $Y^2{\bar Y}^2$  whose K\"ahler variation \eqref{supKahler} should be in turn compensated by the variation of $e^{-\frac {2}3 K}$. We thus re-write \eqref{hatF} in the following form
\be\label{F}
\hat{\mathcal F}_{\Phi,\bar\Phi,V,\varphi}(\xi)=Y^2{\bar Y}^2\,e^{-\frac {2}3 K(\Phi,\bar\Phi,V)}\,{\mathcal F}[\Phi(z(\xi)),\bar\Phi(z(\xi)),V(z(\xi)),\varphi(\xi)]\,,
\ee
where $\mathcal F$ is a gauge-invariant function of bulk and worldvolume fields.

We are now in a position to compare the coupling of the goldstino brane to supergravity and matter with that of the chiral nilpotent superfield of Section \ref{sugra+S}. A straightforward way to do this is to impose the static gauge $x^m=\xi^i \delta_i^m$, the unitary gauge \eqref{chi=0} and gauge fix the super-Weyl invariance by choosing $Y$ as in \eqref{EF}. Then the goldstino brane term in the action \eqref{VAmattercoupling} reduces to
\be\label{VAgf}
S_{\rm brane}=-f^2\int {\rm d}^4x\,(\det e){\mathcal F}[\Phi,\bar\Phi,V,\varphi(x)]|_{\theta=\bar\theta=0}\,.
\ee
Comparing \eqref{VAgf} with \eqref{S+US} and \eqref{gV} we see that the two actions produce the same supersymmetry breaking potentials when
\bea\label{F=U}
 & \!\!\!\!\!\!\! f^2 {\mathcal F}[\Phi,\bar\Phi,V,\varphi(x)]|_{\theta=\bar\theta=0}
= \frac{|\frac 14(\bar{\mathcal D}\bar {\mathcal D}-4R)K_s +{m}\, e^{\frac{K}2}\,(W_s+g^s_{AB}\mathbb W^{A\alpha}\mathbb W_\alpha^B)|^2}{ K_{s\bar s}-\frac {1}3 K_s\bar K_{\bar s} }|_{\theta=\bar\theta=0}\,.&
\eea
If the brane worldvolume fields $\varphi(x)$ are switched off, we see from eq. \eqref{F=U} that the two descriptions of the goldstino couplings easily match for properly chosen functions of the bulk fields.

As we have already mentioned, in addition to the constrained goldstino superfield there may be other constrained superfields involved into the construction of effective models. In the goldstino brane formulation there are different ways of including into consideration the other constrained superfields, like \eqref{SSQ} and \eqref{ST}. For instance, one can use constraints similar to \eqref{SSQ} and \eqref{ST} but with the nilpotent goldstino superfield $S$ replaced by $X$ constructed as in  \eqref{XV1} with $\mathcal V$ defined in \eqref{V=VA}.

Or one can include into the function $\mathcal F$ goldstino brane terms which give large masses to certain components of matter superfields and take an infinite mass decoupling limit. For instance, in the case of the inflaton superfield $T$ the corresponding terms in $\mathcal F$ which lead to the constraint $\eqref{ST}$ have the following form
\be\label{FT}
\mathcal F_T(T,\bar T)={c_1}(T-\bar T)^2+c_2 (\mathcal DT)^2+\bar c_2(\bar{\mathcal D}\bar T)^2+c_3 |\mathcal D\mathcal D T|^2,\qquad c_1, |c_2| ,c_3 \rightarrow \infty\,.
\ee

One can also include the contribution to the effective theory of  brane worldvolume fields. We we will consider examples of these in the forthcoming sections.  Their effect is similar to that of constrained matter superfields in accordance with a recent consideration of \cite{Vercnocke:2016fbt,Kallosh:2016aep}. Before passing to the discussion of these couplings, let us indicate one more feature of the goldstino brane.

 \section{Brane contribution to the gravitino mass and special minimal supergravity}\label{gm}
 The brane may also naturally produce supersymmetry breaking gravitino mass terms in a form similar to the one appeared  in \cite{Volkov:1973jd}
\be\label{deltam}
 \frac{\Delta m}{4\ii\kappa^2} \int {\rm d}^4\xi\, \varepsilon^{ijkl}\,(\mathbb{E}_i\sigma_{ab}\mathbb{E}_j)\, \mathbb{E}^a_k \,\mathbb{E}^b_l+ {\rm c.c.}
\ee
which in the unitary gauge \eqref{chi=0} and the Einstein frame reduces to
\be\label{deltamuni}
-\frac{\Delta m}{2\kappa^2} \int {\rm d}^4x (\det e) \psi_a\sigma^{ab}\psi_b+ {\rm c.c.} \qquad ({\rm at}\,\,\chi=\bar\chi=0,\,\xi^m=x^m).
\ee
With the use of the pullbacks of the compensator superfields and a chiral superpotential the action \eqref{deltam} can be made super-Weyl and K\"ahler invariant
\be\label{deltam1}
\frac{\Delta m}{4\ii\kappa^2} \int {\rm d}^4\xi\, \varepsilon^{ijkl}\,(\hat{\mathbb{E}}_i\sigma_{ab}\,\hat {\mathbb{E}}_j)\, \mathbb{E}^a_k \,\mathbb{E}^b_l\,{\bar Y}^3(z(\xi))\,\bar {\mathcal W}(\Phi(z(\xi))+ {\rm c.c.}\,
\ee
where
\be\label{hatE}
\hat{\mathbb{E}}^\alpha_i=\mathbb E_{i}^\alpha-\frac{\ii}{6}\mathbb E_{i}^a\sigma^{\alpha\dot\alpha}_a\,\bar{\mathcal D}_{\dot\alpha}\log\Big (\bar Y^3\bar {\mathcal W}\Big)\,.
\ee
We stress that in general $\mathcal W$ does not need to coincide with the superpotential defining the F-term in the supersymmetric matter sector of the Lagrangian. It should just transform in the same way under K\"ahler transformations or, in other words, it must be a section of the same line bundle.

The presence of terms like \eqref{deltam} in the effective action may be of phenomenological relevance. As far as we know, manifestly supersymmetric terms like this have not been constructed so far  by using constrained superfields.\footnote{We thank F.~Farakos for discussions of this issue.} A possibility of getting such terms in the constrained superfield approach is as follows.

 We first notice that the integrand of \eqref{deltamuni} is the worldvolume pullback of a part of the four-superform
 \be\label{Omega4}
\Omega_4=\frac {\ii}{4}E^\alpha\wedge (\sigma_{ba})_{\alpha\beta}E^\beta\wedge E^a\wedge E^b+\frac1{32}R(z)\,E^a\wedge E^b\wedge E^c\wedge E^d\varepsilon_{dcba},
\ee
which is closed when one takes into account the $N=1$, $D=4$ supergravity constraints \cite{Binetruy:1996xw,Ovrut:1997ur,Kuzenko:2005wh,Bandos:2011fw,Bandos:2012gz})
\be\label{dO4=0}
\d\Omega_4=0\,.
\ee
This gives rise to a so called {\it 3-form} or {\it special minimal} off-shell formulation of $N=1$, $D=4$ supergravity \cite{Grisaru:1981xm,Ovrut:1997ur,Kuzenko:2005wh,Bandos:2011fw,Bandos:2012gz} tracing roots back to \cite{Stelle:1978ye} and \cite{Ogievetsky:1978mt,Ogievetsky:1980qp}. In this formulation a real part of \eqref{Omega4} is assumed to be an exact form
$$
\Omega_4+\bar \Omega_4=\d C_3.
$$
where $C_3$ is a real  three-superform whose variation under the supersymmetry transformations is a total derivative. Hence
\be\label{A=B}
\frac1{32}\big[R(z)+\bar R(z)\big]\,E^a\wedge E^b\wedge E^c\wedge E^d\varepsilon_{dcba} =\d C_3-\frac{\ii}{4}\left(E^\alpha\wedge (\sigma_{ba})_{\alpha\beta}E^\beta\wedge E^a\wedge E^b- \text{c.c.}\right)\,,
\ee
From eq. \eqref{A=B} it follows that in the special minimal supergravity the real part of the scalar curvature superfield $R(z)$ takes the form
\be\label{RS=dO}
R(z)+\bar R(z) =-\frac {2}{9} \varepsilon^{abcd} E_a^{ M}E_b^{ N} E_c^{P}E_d^{Q}\partial_{[M}C_{{NPQ})}(z),
\ee
where $E_a^{M}(z)$ are components of the vielbein inverse of $E^{ A}_{M}(z)$. In particular, the real part of the auxiliary field $R(x)$ which is the leading component of \eqref{RS=dO} is expressed in terms of the stress tensor of $C_3(x)$ and the gravitino mass terms
\be\label{R=dO}
R(x)+\bar R(x)=- \frac 2{9\, e}\varepsilon^{klmn}\partial_{k}C_{lmn}(x)+\frac 23 (\psi_a\sigma^{ab}\psi_b+{\bar\psi}_a{\tilde\sigma}^{ab}{\bar\psi}_b)
\ee
($e= \det e_\mu^a(x)$).
So in the special three-form supergravity formulation we can replace \eqref{deltam} with
\be\label{deltamR}
 -\frac{3\Delta m}{4\kappa^2} \int \d^4\xi\, \big[R(z(\xi))+\bar R(z(\xi))\big] \det {\mathbb E}.
\ee
Note that in flat superspace the term  \eqref{deltamR} vanishes, and so does    \eqref{deltam} which  becomes an integral of a total derivative \cite{Bandos:2010yy}, hence the gravitino mass term \eqref{deltam} does not have rigid supersymmetry counterparts.

We can now lift \eqref{deltamR} to the complete superspace integral writing down
\be\label{deltamRs}
 -\frac{3\Delta m}{4\kappa^2} \int \d^8z\, {\rm Ber}\,E(z) \big[R(z)+\bar R(z)\big] \mathcal V(z),
\ee
where $\mathcal V(z)$ was defined in \eqref{V=VA}.  In view of  the relations between different constrained superfields discussed in Section \ref{CSF},  $\mathcal V(z)$ can be replaced e.g. with its solution in terms of the chiral superfields $X\bar X$ or the complex linear superfields $\Sigma\bar\Sigma$, or with the bilinear of the nilpotent superfields $S\bar S$ \eqref{S}.

The super-Weyl invariant form of \eqref{deltamRs} is
\be\label{deltamRsY}
\frac{3\Delta m}{16\kappa^2} \int \d^8z\, {\rm Ber}\,E(z) \, P^{2} \Big[Y^{-2}\Big(\bar{{\cal D}}\bar{{\cal D}}-4R(z)\Big)\bar{Y}+\bar{Y}^{-2}\Big({\cal D}{\cal D}-4\bar R(z)\Big)Y\Big] \mathcal V(z),
\ee
where $Y$ are the compensators of special minimal supergravity \cite{Kuzenko:2005wh}  constructed as the chiral projection of a real pre-potential $P$
\be\label{Y=bDbD-RP}
Y^3=(\bar{{\cal D}}\bar{{\cal D}}-4R) P\; , \qquad \bar{Y}{}^3=({\cal D}{\cal D}-4\bar{R}) P\; , \qquad P=\bar{P}\; .
\ee
The scaling properties of $Y$ under the Weyl transformations \eqref{superWeyl} are determined by the following transformations of $P$ \cite{Kuzenko:2005wh}
\be\label{P-Weyl}
P\mapsto P e^{-2(\Upsilon+ \bar{\Upsilon}) }
\;  \qquad
\ee
and of the chiral projector
\be\label{DD-Weyl}
(\bar{{\cal D}}\bar{{\cal D}}-4R) \ldots \quad \to \quad
e^{-4\Upsilon }(\bar{{\cal D}}\bar{{\cal D}}-4R)  e^{ 2\bar{\Upsilon} } \ldots .  \qquad
\ee
The power of the pre-potential $P$ in  \eqref{deltamRsY} is fixed by the super-Weyl weight of the Volkov-Akulov brane superfield
${\cal V}$ \eqref{V=VAcov},
\be\label{cV-Weyl}
{\cal V}\quad \to \quad  {\cal V} e^{2(\Upsilon+ \bar{\Upsilon}) }
\; . \qquad
\ee
By appropriately choosing the power of $P$ one can consider the action  \eqref{deltamRsY} with the real superfield $\mathcal V$ of an arbitrary Weyl weight.

We have thus constructed the Weyl invariant spontaneous supersymmetry breaking contribution to the gravitino mass with the use of a constrained superfield. This construction requires to couple the goldstino superfield to the special minimal supergravity. Further generalization of \eqref{deltamRsY} which would include K\"ahler-invariant matter coupling as in \eqref{deltam1} encounters an obstacle related to the fact that in the minimal special supergravity it is not directly possible to assume that the compensator $Y$ transforms under the K\"ahler transformations, as in \eqref{supKahler}, since the expression of $Y$ in terms of $P$ \eqref{Y=bDbD-RP} does not allow this. So in this version of supergravity matter coupling is more restrictive than in the old minimal one. Further discussion of this issue is beyond the scope of this paper.

To summarize, the goldstino brane can provide the contribution to the gravitino mass in matter-coupled old minimal supergravity, while the constrained superfield counterpart of this term requires coupling to the special minimal supergravity whose interactions with matter superfields are quite restricted. This example may imply that the  goldstino brane and the constrained superfield descriptions are actually not completely equivalent.

\section{Adding matter fields on the brane worldvolume}\label{wv}
In the context of four-dimensional model building one can consider rather general couplings of brane worldvolume fields to goldstino and bulk fields.

The worldvolume fields to which the goldstino may be coupled in the supersymmetry invariant way \cite{Volkov:1972jx,Volkov:1973ix,Akulov:1974xz} transform non-linearly under spontaneously broken supersymmetries and hence do not form supermultiplets. They can however be promoted to constrained superfields following the general procedure of \cite{Ivanov:1977my,Ivanov:1978mx} recently applied to an anti-D3-brane case in \cite{Vercnocke:2016fbt,Kallosh:2016aep}.

The form of the supersymmetry variations of the worldvolume matter fields can be deduced as follows. Consider, for simplicity, a single worldvolume scalar field $\varphi(\xi)$ which is \emph{a priori} inert under the bulk supersymmetry transformations. The only requirement is that it couples to the  goldstino in a worldvolume diffeomorphism invariant way, e.g.
\be\label{Svarphi}
S_\phi=-\frac 12 \int {\rm d}^4\xi \sqrt{-g}\,(g^{ij}\partial_i\varphi\partial_j\varphi+V(\varphi))\,
\ee
where $g_{ij}=\eta_{ab}\mathbb E^a_i\mathbb E^b_j$ is the induced worldvolume metric and $\mathbb E^a_i$ was defined in \eqref{Eapull}. Note that, by construction, this action is also manifestly invariant under the bulk superdiffeomorphisms.

If we impose the static gauge \eqref{staticgauge}, the symmetry which preserves this gauge is a combination of the worldvolume diffeomorphism and the target-space superdiffeomorphisms $\delta x^m=\delta^m_i\delta \xi^i$. Thus, in the static gauge the worldvolume fields undergo the following target-space supersymmetry transformations
\be\label{deltavp}
\delta_\epsilon\varphi(x)=-\delta_\epsilon x^m\partial_m\varphi(x)\,,
\ee
where in the Wess-Zumino gauge (see e.g. \cite{Bandos:2015xnf})
\be\label{deltax}
\delta_\epsilon x^m=\ii f^{-1}\left(\chi \sigma^n\bar\epsilon-  \epsilon\sigma^n\bar\chi \right)
 \left(\delta_n{}^m - \ii f^{-1} \chi\sigma^m\bar\psi_n +  \ii f^{-1}\psi_n \sigma^m\bar\chi \right)
   + {\cal O}(\chi^3) \; .
\ee

A class of effective four-dimensional models with matter fields $\varphi(\xi)$ can be narrowed if the goldstino brane originates from an (anti) D-branes of string theory propagating in a compactified ten-dimensional space-time. The form of the D-brane action (which is of a Dirac-Born-Infeld type) and hence the structure of the terms involving the worldvolume fields are fixed by the worldvolume and target-space symmetries \cite{Cederwall:1996pv,Cederwall:1996ri,Aganagic:1996pe,Bergshoeff:1996tu}. The relation between the Volkov-Akulov model and superbrane effective actions has been comprehensively discussed in the literature, see e.g. \cite{Hughes:1986dn,Kallosh:1997aw,Bergshoeff:2015jxa,Dasgupta:2016prs,Vercnocke:2016fbt,Kallosh:2016aep} to which we address the reader for further details and references. We only note that, in the case of an anti-D3-brane ($\overline{\rm D3}$-brane), these fields may include six scalar fields $y^{p}(\xi)$ $(p=1,\ldots,6)$ associated with brane embedding coordinates of the  compactified internal space, a $U(1)$ Born-Infeld gauge field $A_i(\xi)$ and extra 3 fermionic ($D=4$ spinor) fields $\psi^I(\xi)$ (I=1,2,3). When in a given $D=10$ background the kappa-symmetry of the D3-brane action is appropriately fixed and the static gauge \eqref{staticgauge} is imposed, the fields $y^{p}(x)$, $A_m(x)$ and $\psi^I(x)$ transform under spontaneously broken $\mathcal N=1$, $D=4$ supersymmetry similar to \eqref{deltavp} (see e.g. \cite{Bagger:1996wp,Vercnocke:2016fbt,Kallosh:2016aep})
\be
\delta_{\epsilon}y^{p}=-\delta_{\epsilon} x^m\del_m y^{p}\,,\quad \delta_{\epsilon}A_m=-\delta_{\epsilon} x^n F_{nm}-\partial_m(\delta_\epsilon x^nA_n)\,,\quad \delta_{\epsilon}\psi^I=-\delta_{\epsilon} x^\mu\del_\mu \psi^I.
\ee
The presence of fluxes and orientifold planes, e.g.\ when the $\overline{\rm D3}$-brane sits in a strongly warped region of a type IIB flux compactification discussed in the next section, can remove part of the $\overline{\rm D3}$-brane fields from the low-energy spectrum (see e.g. \cite{Bergshoeff:2005yp,Bergshoeff:2015jxa,Kallosh:2016aep} and references therein for a detailed discussion of this case) and significantly affect the contribution of the $\overline{\rm D3}$-brane to the effective theory.

Let us consider a general situation (not necessarily associated with anti-D-branes in string compactifications)  in which worldvolume bosons $y^p(\xi)$, fermions $\psi^I$ and a $U(1)$ gauge field $A_i(\xi)$    are part of the spectrum of the low-energy effective  theory. Then, to the second order in derivatives, the function $\hat{\mathcal F}$ in the goldstino brane term of the action \eqref{VAmattercoupling} may have the following generic form
\be\label{Fw}
\begin{aligned}
 \hat{\mathcal F}=&
\,U[\Phi(z(\xi)),\bar\Phi(z(\xi)),V(z(\xi)),y(\xi), \psi(\xi)]\\
 &+\frac1{4} {\rm g}[\Phi(z(\xi)),\bar\Phi(z(\xi)),V(z(\xi)),y(\xi),\psi(\xi)]\,g^{ik}(\xi)g^{jl}(\xi) F_{ij}(\xi)F_{kl}(\xi)\\
 &+C[\Phi(z(\xi)),\bar\Phi(z(\xi)),V(z(\xi)),y(\xi),\psi(\xi)]F_{ij}F^{*\,ij}&\\
 &+\frac12G_{pq}[\Phi(z(\xi)),\bar\Phi(z(\xi)),V(z(\xi)),y(\xi),\psi(\xi)]\,g^{ij}(\xi) \del_iy^p (\xi) \del_jy^q(\xi)\,\\
& +\frac 12 G_{IJ}[\Phi(z(\xi)),\bar\Phi(z(\xi)),V(z(\xi)),y(\xi), \psi(\xi)]\psi^I\slashed \nabla \psi^J\,,&
 \end{aligned}
\ee
where $\nabla$ is a covariant derivative constructed with an induced spin connection, $F_{ij}=\partial_iA_j-\partial_jA_i$, $F^{*\,ij}=\frac 1{\sqrt{-g}} \varepsilon^{ijkl}F_{kl}$ and all the superfunctions are gauge invariant. For this choice of $\hat{\mathcal F}$, in the static and unitary gauge the goldstino brane action \eqref{VAgf} takes the form
\bea\label{genL}
S_{\rm brane}&=&-f^2\int \d^4x \,\det e \,\Big[U(\Phi,\bar\Phi,V,y,\psi)+\frac14 {\rm g}(\Phi,\bar\Phi,V,y,\psi)F_{mn}F^{mn}\nonumber\\
&&+C[\Phi,\bar\Phi,V,y,\psi]F_{mn}F^{*\,mn}+\frac12G_{pq}(\Phi,\bar\Phi,V,y,\psi)g^{mn} \del_my^p \partial_ny^q\nonumber\\
&&+\frac 12 G_{IJ}(\Phi,\bar\Phi,V,y, \psi)\,\psi^I\slashed \nabla \psi^J\Big]|_{\chi=\bar\chi=0}.
\eea

This illustrates the generality of the goldstino brane approach to the construction of locally-supersymmetric actions which describe couplings of a supersymmetry breaking sector to a supersymmetric bulk theory. To reduce the range of possible models one should therefore resort to the 4D effective field theory description of concrete phenomenologically relevant scenarios, such as e.g. flux compactifications with D-branes and anti-D-branes in string theory. This has been a subject of a significant interest over the years.

\section{$4D$ effective action for a $\overline{\rm D3}$-brane in flux compactifications of type IIB string theory}\label{aD3}
\label{sec:antibrane}

 Let us briefly review a particular form of \eqref{genL} which corresponds to  well known examples of flux compactifications with branes in string theory, namely the KKL(MM)T models \cite{Kachru:2003aw,Kachru:2003sx} whose key ingredients are $\overline{\rm D3}$-branes.

 In the framework of the simplest  KKL(MM)T setups \cite{Kachru:2003aw,Kachru:2003sx}, it is assumed that there is one $\overline{\rm D3}$-brane confined at the bottom of a strongly warped region \cite{Giddings:2001yu}, for instance the one similar to the Klebanov-Strassler solution \cite{Klebanov:2000hb}. As a further simplification,  it is often assumed that there is just one bulk (universal) K\"ahler modulus plus, possibly, the moduli describing the position of (supersymmetric) mobile D3-branes. The supersymmetrization of this setting has been discussed with the use of constrained superfields (see e.g. \cite{Vercnocke:2016fbt,Kallosh:2016aep} and references therein). Here we would like to revisit this problem and show how the goldstino brane allows for a natural solution of it in the framework of Section \ref{wv}. In passing,  by using the warped  effective supergravity of \cite{Martucci:2014ska,Martucci:2016pzt}, we will rederive and extend the formulas obtained in  \cite{Kachru:2003aw,Kachru:2003sx} for describing the contribution of the $\overline{\rm D3}$-brane to the potential.

\subsection{Type IIB compactification background}
In order to identify the contribution of the $\overline{\rm D3}$-brane to the $4D$ effective theory in the KKL(MM)T scenarious \cite{Kachru:2003aw,Kachru:2003sx}, we first briefly recall the structure of the warped compactifications in type IIB string theory described in \cite{Giddings:2001yu} (partially) following the notation of \cite{Martucci:2014ska,Martucci:2016pzt}. The ten-dimensional Einstein-frame metric of a tree-level {\em vacuum} is
\be\label{10dmetric}
\d \hat s^2_{10}=\ell_{\rm s}^2\,M_P^2 e^{2A(y)}|Y|^2\d s^2_4(x)+\ell_{\rm s}^2\, e^{-2A(y)}\d s^2_{X_6}(y)
\ee
where $l_s$ is a string length scale, $M_P=\frac 1\kappa$ is the effective $4D$ Plank mass, $\d s^2_4=  g_{(4) mn}(x) \d x^m \d x^m$ is a $4D$ metric around which the $4D$ effective field theory is constructed, $e^{2A(y)}$ is the warping factor which non-trivially depends on the coordinates $y^p$  of an internal space endowed with a K\"ahler metric $\d s^2_{X_6}=g_{pq}(y)\d y^p\d y^q $ ($p,q,\ldots=1,\ldots, 6$).
In \eqref{10dmetric} we have factorised the string-length dependence $\ell_{\rm s}^2=(2\pi)^2\alpha'$ in order to work in natural string units along the internal space.
Finally $Y$ is associated with the lowest component of the auxiliary compensator superfield introduced in Section \ref{sec:coupling} to single out the $D=4$ Einstein frame by fixing a specific value of $Y$ determined by the K\"ahler potential as in \eqref{EF}.

The general form of the warping is
\be\label{wsol}
e^{-4A}=a +e^{-4A_0(y)} \quad~~~~\text{with}\quad  e^{-4A_0(y)}=\frac{1}{\ell^4_{\rm s}}\int_{X_6} G(y;y')Q_6(y')\,,
\ee
where
 $a$ is an arbitrary parameter, the so called {\em universal modulus} which in the absence of $e^{-4A_0(y)}$ is associated with the overall volume of the internal space $X_6$,
\ $G(y;y')$ is a Green's function associated with the internal space metric and  $Q_6(y)$ is the six-form D3-charge density which encodes the contribution of $N_{\rm D3}$ mobile D3-branes, $H_3$ and $F_3$ fluxes, O3-planes  and, possibly, other localised sources:
\be\label{D3charge}
Q_6=F_3\wedge H_3+\ell_{\rm s}^4\sum_{I\in\text{D3's}}\delta^6_I-\frac14\ell_{\rm s}^4\sum_{O\in\text{O3's}}\delta^6_O+\ldots\,.
\ee
In this background there is also a self-dual RR five-form flux
\be\label{F5}
F_5=\ell^4_{\rm s}\,M_P^4\,|Y|^4\d{\rm vol}_4\wedge \d e^{4A}+\ell^4_{\rm s}*_6\d e^{-4A}\,,
\ee
For simplicity, we assume that the axion-dilaton is constant, i.e.
\be\label{ad}
\tau\equiv C_0+ \ii e^{-\phi}=c_0+\frac\ii{g_s}.
\ee
Furthermore, we assume that its value as well as the complex structure moduli are fixed dynamically by the fluxes.

The bosonic fields of the effective four-dimensional theory describing excitations around the vacuum under consideration include, in addition to the four-dimensional metric $g_{(4)mn}$, a set of complex fields $\rho^a(x)$ which parametrize the  $X_6$ K\"ahler structure and $C_4$ moduli, and complex fields $z^i_I(x)$ that describe the position of the mobile D3-branes (in some local complex coordinate $z^i$ in the internal space $X_6$ and $I=1,\ldots, N_{\rm D3}$). The fields $\rho^a$ and $z^i_I$ are the lowest components of chiral superfields, which we denote with the same symbols. There may be additional fields contributing to the effective four-dimensional theory but for simplicity we neglect them.

The four-dimensional effective field theory is specified by a superpotential and a K\"ahler potential for these fields. In \cite{Martucci:2014ska} it is shown that the K\"ahler potential $K(\rho,\bar\rho,z,\bar z)$ is implicitly defined by the following simple formula
\be\label{wK}
K=-3\log a\,,
\ee
where the universal modulus $a$ is considered as a function of the chiral moduli $\rho$ and $z^i_I$.   To describe the effective theory in the $D=4$ Einstein frame in what follows we set
\be\label{confcomp}
Y= e^{\frac {K}6}\,,
\ee
which is just the bosonic part of the full superfield gauge-fixing condition (\ref{EF}). Notice that the specific choice (\ref{wK}) explicitly breaks the K\"ahler invariance of the effective theory.

\subsection{$\overline{\rm D3}$-brane contribution: bosonic fields}

Let us now add an anti-${\rm D3}$-brane to the above configuration. As in \cite{Kachru:2003aw,Kachru:2003sx} we consider the $\overline{\rm D3}$-brane as a probe, i.e.\ we work in the approximation in which its backreation on the background is neglected. This means, in particular, that  we neglect the (negative) contribution of the $\overline{\rm D3}$-brane to the charge density (\ref{D3charge}), keeping in it only supersymmetry preserving sources. In fact, since $X_6$ is compact, this contribution would modify the tadpole condition $\int_{X_6} Q_6=0$, which gives the global consistency of the configuration. Hence, the ten-dimensional backreaction of the $\overline{\rm D3}$-brane should be taken  into account along the lines of \cite{Michel:2014lva,Polchinski:2015bea},  at least to accommodate for such an effect, but we will not try to do it here\footnote{There exists the so-called `complete but gauge fixed' Lagrangian description of supergravity-superbrane interacting systems \cite{Bandos:2001jx,Bandos:2002kk,Bandos:2005ww}, which  can also be applied to study the $\overline{\rm D3}$-brane backreaction. }.

The potential felt by the $\overline{\rm D3}$-brane in the background under consideration can be extracted from the standard bosonic D3-brane effective action in the $D=10$ Einstein frame
\be\label{antiD3action}
S_{\overline{\rm D3}}=-\frac{2\pi}{\ell^4_{\rm s}}\int \d^4\xi \sqrt{-\det\left[\hat g_{ij}+e^{-\frac\phi2}\calf_{ij}\right]}-\frac{2\pi}{\ell^4_{\rm s}}\int C\wedge e^{\mathcal F}
\ee
where
$\hat g_{ij}= \partial_i x^m \partial_j x^n g_{mn}(x,y) + \partial_i y^p \partial_j y^q g_{pq}(y)$ is
the induced metric,
$\calf_{ij}\equiv \frac{\ell^2_{\rm s}}{2\pi}F_{ij}- B_{ij}$ with $F_{ij}=2\partial_{[i}A_{j]}(\xi)$ being the  $\overline{\rm D3}$-brane Born-Infeld field and $B_{ij}$ being the pull-back of the NS-NS two-form. The second (Wess-Zumino) term describes the couplings of the $\overline{\rm D3}$-brane to Ramond-Ramond potentials $C_0$, $C_2$ and $C_4$, and
$\int C\wedge e^{\mathcal F}= \int C_4 + \int {\mathcal F}\wedge C_2 + \frac 1 2\int C_0\,  {\mathcal F}\wedge {\mathcal F}$. In our conventions the overall minus sign of the WZ terms signals  that we deal with the anti-D3-brane rather than the D3-brane.

Upon fixing the static gauge $\xi^i=\delta^i_m x^m$, from (\ref{antiD3action}) one can read that in the bulk under consideration -- see \eqref{10dmetric}, \eqref{wsol}, \eqref{F5}, \eqref{wK} and \eqref{confcomp} --  the $\overline{\rm D3}$-brane at a point $y^p$ feels the  effective potential
\be\label{antiD3pot1}
U_{\overline{\rm D3}}(y)=4\pi M_P^4 |Y|^4 e^{4A(y)}=4\pi  M_P^4  \frac{ e^{\frac{2 K}3}}{e^{-\frac {K}3 }+e^{-4A_0(y)}}\,.
\ee
Notice that the DBI and the WZ part of the $\overline{\rm D3}$-brane action give equal contributions into $U_{\overline{\rm D3}}$. These contributions would cancel each other in the $D3$-brane case in which the sign of the WZ term is opposite.
For fixed $\rho^a$ and $z^i_I$ moduli, this potential attracts the $\overline{\rm D3}$-brane to the regions of the internal space in which $e^{4A(y)}$ has a minimum or, equivalently, where  $e^{-4A_0(y)}$ has a maximum.

Suppose that $e^{-4A_0(y)}$ is maximised at an isolated point
$y^p_{\overline{\rm D3}}$ and that around it the potential is very steep.
We can then integrate out the field describing the position of the $\overline{\rm D3}$-brane and get the following contribution to the low-energy $4D$ effective action
\be\label{antiD3pot2}
-\int \d^4 x  \sqrt{-\det g_4}\,U_{\overline{\rm D3}}=- \frac {4\pi} {\kappa^4} \int \d^4 x  \sqrt{-\det g_4}\,\frac{e^{K}}{1+e^{\frac { K}3  }e^{-4A_0(y_{\overline{\rm D3}})}}.
\ee
The potential (\ref{antiD3pot2}) agrees with those of \cite{Kachru:2003aw,Kachru:2003sx} in particular simplifying limits. This is discussed in Appendix \ref{app:antiD3}. Having integrated out the $\overline{\rm D3}$ embedding fields $y^p(x)$, the only low-energy bosonic world-volume field which remains  is the gauge field $A_i$ (unless it is removed by an orientifold projection). Its leading two-derivative contributions to the effective action are obtained by expanding (\ref{antiD3action}) (in the static gauge)
\be\label{antiD3kin1}
S_{U(1)}=-\frac{1}{8\pi g_s}\int \d^4 x  \sqrt{-\det g_4}\, F_{mn}F^{mn}-\frac{c_0}{4\pi}\int \, F\wedge F\,.
\ee
Note that in the simplest case of constant axion $c_0$ the second term is purely topological.

One can also consider situations in which the minimum of (\ref{antiD3pot1}) is
degenerate, as for instance  along the $S^3$ at the bottom of the Klebanov-Strassler  throat \cite{Klebanov:2000hb}. Then the fluctuations of the $\overline{\rm D3}$-brane  $\varphi^A(x)$ (${\mathcal A}=1,2,3$) along the  $S^3$ are massless dynamical fields in the low-energy effective action. At the two-derivative level they are described by the action
\be\label{antiD3kin2}
S_{\varphi^A}=-\pi M^2_{\rm P}\int \d^4 x \sqrt{-\det g_4}\, e^{\frac {K}3} e^{2A} \calg_{\mathcal{AB}}(\varphi)g^{mn}\del_m\varphi^{\mathcal A}\del_n\varphi^{\mathcal B}\,,
\ee
Comparing \eqref{genL} with (\ref{antiD3pot2}), \eqref{antiD3kin1} and \eqref{antiD3kin2} we see that the latter three actions are particular examples of the generic goldstino brane construction.

\subsection{$\overline{\rm D3}$-brane contribution: goldstino and other fermions}
To include into consideration the $\overline{\rm D3}$-brane fermionic modes one should promote all the background fields of the action \eqref{antiD3action} to supefields in $D=10$ type IIB superspace \cite{Cederwall:1996pv,Cederwall:1996ri,Aganagic:1996pe,Bergshoeff:1996tu}. Then, as usual, the fermionic worldvolume fields appear as brane embedding coordiantes along the Grassmann-odd directions of the superspace (see the corresponding discussion in Section \ref{VA}). The explicit form of the fermionic part of the D-brane actions in generic $D=10$ backgrounds is known only to the second order in the fermionic fields \cite{Marolf:2003vf}. In the case of our interest the quadratic fermionic action has the following schematic form
\be\label{fermiaction}
S_{\rm fermi}=-\frac{\ii\pi}{\ell^4_{\rm s}}\int \d^4\xi \sqrt{-\det\left[\hat g_{ij}+e^{-\frac\phi2}\calf_{ij}\right]}\,\bar \Theta \slashed {\mathcal D}\Theta\,,
\ee
where (upon gauge fixing $\kappa$-symmetry) $\Theta(\xi)$ describes 16 dynamical fermionic degrees of freedom of the $\overline{\rm D3}$-brane and $\slashed{\mathcal D}$ is a generalized Dirac operator whose form depends on the pull-back of the background fields and the worldvolume BI field $F_{ij}$.

By dimensionally reducing this action in the background of our interest to four dimensions, one finds that the supersymmetric bulk fluxes generate masses for 12 of the 16 dynamical fermions  $\Theta(\xi)$, as discussed in detail in \cite{Bergshoeff:2015jxa}, leaving only four massless real fermionic fields, which are identified with the four-dimensional goldstino.
One can see that in the static gauge $\xi^i=\delta^i_m x^m $ the action \eqref{fermiaction} is a particular case of the fermionic part of the generic action (\ref{genL}).

This illustrates how the general approach for coupling the goldstino to supergravity and to other supersymmetric or non-supersymmetric matter discussed in the present paper naturally includes the four-dimensional effective models obtained by the dimensional reduction of D-brane actions in string theory. Notice that this  matching can be in principle extended to all higher order terms in the expansion of the D-brane action, by appropriately generalising the four-dimensional brane action (\ref{genL}) to the Dirac-Born-Infeld-like one, along the general lines described in the previous sections.

\section{Conclusion}
We have shown that different ways of describing the goldstino in $\mathcal N=1,$ $D=4$ supersymmetric theories coupled to supergravity, either in terms of the constrained superfields or as the original Volkov-Akulov brane-like construction, are equivalent to each other and lead to similar very general effective field-theoretical models.
The choice of one or another formulation may depend on the choice of the setup for model building.
In particular, as we have demonstrated, the use of the goldstino brane provides a more geometrically intuitive way of constructing the couplings of the goldstino to matter supermultiplets, supergravity and single (`non-supersymmetric') matter fields, which is related in a more direct way to stringy phenomenological model building with flux compactification and branes.

We have shown that goldstino brane couplings to supergravity naturally produce an additional contribution to the gravitino mass which may be relevant in concrete phenomenological setups. On the other hand, getting such terms in the constrained superfield approaches and especially coupling them to matter is less straightforward. It would be of interest to understand whether and how such terms might arise from the dimensional reduction of a D-brane action in ten-dimensional superbackground.

\section*{Acknowledgements}
This work was initiated with an enthusiastic participation of Mario Tonin who, to our great sorrow, left us on April 13, 2016.
S.M.K. acknowledges the generous hospitality of
the INFN, Sezione di Padova, and of the
Max Planck Institute for Gravitational Physics (Albert Einstein Institute), Potsdam.
S.M.K. is grateful to Ian McArthur and Simon Tyler for useful discussions. L.M. and D.S. are thankful to Gianguido Dall'Agata and Fotis Farakos for useful discussions and comments.  The work of I.B.~was partially   supported by the Spanish MINECO grant FPA2012-35043-C02-01,  partially financed  with FEDER funds,  by the Basque Government research group grant ITT559-10 and by the Basque Country University program UFI 11/55. The work of S.M.K. and D.S. is supported in part by the Australian Research Council, project No. DP160103633. The work of L.M.\ was partially supported by the Padua University
Project CPDA144437. The work of D.S.~was partially supported by the Russian Science Foundation grant 14-42-00047 in association with Lebedev Physical Institute.


\begin{appendix}

\section{Classical consistency of the unitary gauge}\label{app:unitarygauge}

Let us show that in any generic matter-coupled supergravity in which spontaneous breaking of local supersymmetry is manifested by the presence of goldstini (as e.g. the action \eqref{VAmattercoupling}), the equations of motion of the latter are not independent, but are consequences of equations of motion of all the physical fields which couple to the goldstini. This property manifests a well known generic fact that the goldstone fields in the theories with spontaneously broken local symmetries are completely auxiliary Stueckelberg-like fields and can be removed by fixing a so called unitary gauge.

The proof is as follows. Let $S_0(\varphi)$ be an action for a set of fields possessing a local symmetry. In the case of our interest we deal with local supersymmetry and $\varphi$ stands for the fields of the supergravity multiplet and all possible physical matter fields. In particular, the action $S_0(\varphi)$ is invariant under the infinitesimal transformations of the fields $\varphi$ with local symmetry parameters $\epsilon(x)$ which have the following generic form
\begin{equation}\label{deltavarphi}
\delta_\epsilon\varphi=\epsilon(x) a(\varphi)+\mathcal D\epsilon(x),
\end{equation}
where $a(\varphi)$ is a field-dependent function, while a covariant derivative $\mathcal D$ of the symmetry parameters appears in the symmetry transformations of the gauge fields, like gravitino.

The invariance of the action implies that (schematically)
\begin{equation}\label{deltaS0}
\delta_\epsilon S_0=\int \delta_\epsilon\varphi(x)\frac{\delta S_0}{\delta\varphi(x)}=\int\epsilon(x)\left(a(\varphi)\frac{\delta S_0}{\delta\varphi}-\mathcal D\frac{\delta S_0}{\delta\varphi}\right)=0\,,
\end{equation}
where (modulo total derivatives) $\frac{\delta S_0}{\delta\varphi}$ stand for the (left-hand sides of the) equations of motion of the fields $\varphi$. Since the parameters $\epsilon(x)$ are arbitrary the invariance of the action implies that the following combinations of the field equations are identically zero (without the use of the equations of motion themselves)
\begin{equation}\label{Noetheri}
a(\varphi)\frac{\delta S_0}{\delta\varphi}-\mathcal D\frac{\delta S_0}{\delta\varphi}\equiv 0\,.
\end{equation}
These are the well-known Noether identities whose number is equal to the number of the local symmetry parameters, the property that constitutes the second Noether theorem.

Let us now add to the action $S_0$ a spontaneous symmetry breaking term $S_1(\chi,\varphi)$ containing a generic coupling of physical fields to goldstone fields $\chi$
\be\label{S0+S1}
S=S_0+S_1\,.
\ee
By construction $S_1(\chi,\varphi)$ is invariant under the local symmetry transformations \eqref{deltavarphi} accompanied by non-linear transformations of the goldstone fields encoded in the function $b(\chi,\varphi)$
\be\label{deltachi}
\delta \chi=\epsilon(x)(1+b(\chi,\varphi)).
\ee
Namely,
\be\label{deltaS1}
\begin{aligned}
\delta_\epsilon S_1&=\int \left[\delta_\epsilon\varphi(x)\frac{\delta S_1}{\delta\varphi(x)}+\delta_\epsilon\chi(x)\frac{\delta S_1}{\delta\chi(x)}\right]\\
&=\int\epsilon(x)\left[a(\varphi)\frac{\delta S_1}{\delta\varphi}-\mathcal D\frac{\delta S_1}{\delta\varphi}+(1+b (\chi,\varphi))\frac{\delta S_1}{\delta\chi}\right]=0\,.
\end{aligned}
\ee
Notice that $\frac{\delta S_1}{\delta\chi}=0$ gives the goldstone field equations of motion. On the other hand, from (\ref{deltaS1}) we see that the invariance of $S_1$ requires that the following equality holds off the mass shell
\be\label{chiphi}
\frac{\delta S_1}{\delta\chi}=-\frac{1}{1+b(\chi,\varphi)}\left(a(\varphi)\frac{\delta S_1}{\delta\varphi}-\mathcal D\frac{\delta S_1}{\delta\varphi}\right)\,.
\ee
Now, the equations of motion of the physical fields get contributions from $S_1$ and take the form
\be\label{eomvarphi}
\frac{\delta S_0}{\delta\varphi}=-\frac{\delta S_1}{\delta\varphi}\,,
\ee
while the Noether identities \eqref{Noetheri} do not change and still hold. Comparing \eqref{Noetheri} and \eqref{eomvarphi} we see that on the mass shell, for consistency, the following combination of the terms on the right-hand side of \eqref{eomvarphi} should be equal to zero on its own
\be\label{S10}
a(\varphi)\frac{\delta S_1}{\delta\varphi}-\mathcal D\frac{\delta S_1}{\delta\varphi}=0\,.
\ee
Comparing \eqref{S10} with \eqref{chiphi} we see that the goldstone field equations are identically satisfied if the physical fields obey their equations of motion \eqref{eomvarphi}. In other words, the goldstone fields do not have independent field equations and hence are completely auxiliary Stueckelberg-like fields. As such, the consistency of \eqref{S10} and \eqref{chiphi} also implies that we can always choose the unitary gauge and put $\chi=0$ directly in the action $S_1$, which then however loses the gauge invariance.

Similar  discussion of the properties of the Lagrangian description of interactions of super-p-branes with dynamical supergravity  provided the basis for the  `complete but gauge fixed description' of these systems \cite{Bandos:2001jx,Bandos:2002kk,Bandos:2005ww}.


\section{The $\overline{\rm D3}$-brane potential: some  explicit cases}
\label{app:antiD3}

Suppose that all the embedding fields  $y^p_{\overline{\rm D3}}$ of the anti-D3-brane are massive and that the bulk sector contains only the universal modulus $a=\frac {2\pi}3(\rho+\bar\rho)$ depending on a single chiral field $\rho$, as in \cite{Kachru:2003aw}. In this case the position of $y^m_{\overline{\rm D3}}$ is determined by the extremisation of  $e^{-4A_0(y_{\overline{\rm D3}})}$, which does not depend on the universal modulus. Hence, at the minimum of the potential  $e^{-4A_0(y_{\overline{\rm D3}})}$ takes a fixed value $e^{-4A^{\rm min}_0}$.
From (\ref{wK}) we get $K=-3\log (\rho+\bar\rho)-3\log\frac{2\pi}{3}$ and then the potential (\ref{antiD3pot2}) becomes
\be
U_{\overline{\rm D3}}(\rho,\bar\rho)=\frac{27 M^4_{\rm P}}{\pi(\rho+\bar\rho)^2}\cdot\frac{1}{2\pi(\rho+\bar\rho)+3e^{-4A^{\rm min}_0}}
\ee
This  potential has two limiting behaviours. If the warping felt by the $\overline{\rm D3}$-brane is weak, so that $e^{-4A^{\rm min}_0}\ll \pi(\rho+\bar\rho)$, we have $U_{\overline{\rm D3}}\sim M^4_{\rm P}(\rho+\bar\rho)^{-3}$ as  in \cite{Kachru:2003aw}.
If the warping is strong, $e^{-4A^{\rm min}_0}\gg \pi(\rho+\bar\rho)$, then we have $U_{\overline{\rm D3}}\sim M^4_{\rm P}e^{4A^{\rm min}_0}(\rho+\bar\rho)^{-2}$ as  in  \cite{Kachru:2003sx}. In particular, one needs strong warping if $\rho$ is only moderately large and one wants a strong suppression of the contribution of the $\overline{\rm D3}$-brane potential. Then we can approximate the potential as
\be
U_{\overline{\rm D3}}(\rho,\bar\rho)\simeq \frac{9 M^4_{\rm P}\,e^{4A^{\rm min}_0}}{\pi(\rho+\bar\rho)^2}\,.
\ee
Still assuming that all the embedding fields  $y^p_{\overline{\rm D3}}$ are massive, we can also  include additional bulk chiral fields $z^i_I$ describing mobile D3-branes. In this case  $\rho+\bar\rho=\frac{3 a}{2\pi}+\sum_I k(z_I,\bar z_I)$, where $k(z,\bar z)$ is the K\"ahler potential of the internal (non-dynamical) metric $\d s^2_{X_6}$. From (\ref{wK}) one then gets
\be
 K=-3\log [\rho+\bar\rho-\sum_I k(z_I,\bar z_I)] -3\log\frac{2\pi}{3}\,,
\ee
as first proposed in  \cite{DeWolfe:2002nn}, which should be used in (\ref{antiD3pot2}) together with
\be
e^{-4A_0(y)}=\sum_I G(y;y_I)+e^{-4\tilde A_0(y)}\,,
\ee
where $e^{-4\tilde A_0(y)}$ comes from the non-dynamical D3-charge entering  $Q_6$ in (\ref{wsol}) and does not depend on any modulus.
As already mentioned  above, the extremisation of (\ref{antiD3pot2}) is equivalent to the extremisation of $e^{-4A_0(y)}$. If the mobile D3-branes are far enough  from the $\overline{\rm D3}$-brane, in a first approximation we can neglect the contribution of  $\sum_I G(y;y_I)$ to $e^{-4A_0(y)}$ in the determination of the minimum $y^p_{\overline{\rm D3}}$, so that
\be
e^{-4A_0(y_{\overline{\rm D3}})}\simeq\sum_I G(y_{\overline{\rm D3}};y_I)+e^{-4\tilde A^{\rm min}_0}
\ee
As above, let us  also assume  that $\overline{\rm D3}$-brane sits at a strongly warped point,  $e^{-4\tilde A^{\rm min}_0}\gg 1$. Then (\ref{antiD3pot2}) gives
\be
U_{\overline{\rm D3}}
(\rho,\bar\rho)
\simeq \frac{9 M^4_{\rm P}\,e^{4\tilde A^{\rm min}_0}}
{\pi\left[\rho+\bar\rho-\sum_I k(z_I,\bar z_I)\right]^2}
\left[1-e^{4\tilde A^{\rm min}_0}\sum_I G(y_{\overline{\rm D3}};y_I)\right]\,.
\ee
In this formula the Green's function encodes the ${\rm D3}$-$\overline{\rm D3}$ Coulomb interaction. This kind of contribution was identified  in \cite{Kachru:2003sx} in the case of conical geometries in which $G(y_{\overline{\rm D3}};y_I)\sim 1/r_I^4$, where $r$ is the radial coordinate.

\end{appendix}



\if{}
\bibliographystyle{abe}
\bibliography{references}{}
\fi

\providecommand{\href}[2]{#2}\begingroup\raggedright\endgroup

\end{document}